\documentclass{article_saj}
\usepackage{graphicx,saj,multicol,subeqnarray}
\usepackage{natbib}
\usepackage{float}
\usepackage{xcolor}
\usepackage{widetext}
\usepackage{cuted}
\usepackage{url}
\usepackage{bm}
\usepackage{tikz} 
\usepackage{pifont} 
\usepackage{amsfonts}
\usepackage{amssymb}
\usepackage{amsmath,upgreek}
\usepackage{titlesec}

\def\point#1{\hbox{\setbox7=\hbox to0.6em{\hfil.\hfil}%
\setbox8=\hbox to0.5em{\hfil$^{#1}$\hfil}%
\box7\kern-0.5em\box8}}

\def\pointmin#1{\hbox{\setbox2=\hbox to0.8em{\hfil.\hfil}%
\setbox3=\hbox to0.6em{\hfil$^{#1}$\hfil}%
\box2\kern-.7em\box3}}

\def\mmm{\pointmin{\mathrm{m}}\kern.15em}

\titlelabel{\thetitle.\quad}
\definecolor{xlinkcolor}{cmyk}{1,0.6,0,0}
\usepackage{amsmath}
\usepackage[bookmarks=false,         
     pdfnewwindow=true,      
     colorlinks=true,    
     linkcolor=xlinkcolor,     
     citecolor=xlinkcolor,     
     filecolor=xlinkcolor,  
     urlcolor=xlinkcolor,      
final=true
]{hyperref}
\usepackage{xspace}
\usepackage[nolist]{acronym}
\usepackage{multicol}
\usepackage{supertabular}
\usepackage{multirow}
\usepackage{comment}

\newcommand{\ergcm}[1]{erg~cm$^{-2}$ s$^{-1}$}

\newcommand{\HI}{H{\sc i}}
\newcommand{\HII}{H{\sc ii}}
\newcommand{\SII}{[S{\sc ii}]}
\newcommand{\OIII}{[O{\sc iii}]}
\newcommand{\Halpha}{H${\alpha}$}

\newcommand{\D}{$^\circ$}

\def\arcmin{\hbox{$^\prime$}}
\def\arcsec{\hbox{$^{\prime\prime}$}}

\newcommand{\farcm}{\mbox{\ensuremath{.\mkern-4mu^\prime}}}
\newcommand{\farcs}{\mbox{\ensuremath{.\!\!^{\prime\prime}}}}
\newcommand{\fdg}{\mbox{\ensuremath{.\!\!^\circ}}}

\begin{acronym}
    \acro{AGN}{Active Galactic Nuclei}
\acro{ALMA}{Atacama Large Millimeter Array}
\acro{ATCA}{Australia Telescope Compact Array}
\acro{ATNF}{Australia Telescope National Facility}
\acro{ATOA}{Australia Telescope Online Archive}
\acro{AT20G}{Australia Telescope 20 GHz Survey}
\acro{ASKAP}{Australian Square Kilometre Array Pathfinder}
\acro{CASS}{CSIRO Astronomy and Space Science}
\acro{CABB}{Compact Array Broadband Backend}
\acro{CC}{core-collapse}
\acro{CSIRO}{Australian Commonwealth Scientific and Industrial Research Organisation}
\acro{CSS}{Compact Steep Spectrum}
\acro{CTA}{Cherenkov Telescope Array}
\acro{DeMCELS}{Dark Energy Camera Magellanic Cloud Emission Line Survey}
\acro{EMU}{Evolutionary Map of the Universe}
\acro{ESP}{Early Science Project}
\acro{FWHM}{Full Width at Half-Maximum power}
\acro{GPS}{Gigahertz Peak Spectrum}
\acro{HFP}{High Frequency Peaker}
\acro{ISM}{interstellar medium}
\acro{ICM}{Inter Cloud Medium}
\acro{IFS}{Intermediate frequencies}
\acro{LMC}{Large Magellanic Cloud}
\acro{MCs}{Small \& Large Magellanic Clouds}
\acro{MCELS}{Magellanic Cloud Emission Line Survey}
\acro{MIRIAD}[{\tt MIRIAD}]{Multichannel Image Reconstruction, Image Analysis and Display}
\acro{MM}{mixed-morphology}
\acro{MOST}{Molonglo Observatory Synthesis Telescope}
\acro{MRC}{Molonglo Reference Catalogue of Radio Sources}
\acro{MWA}{Murchison Widefield Array}
\acro{NRAO}{National Radio Astronomy Observatory}
\acro{NVSS}{NRAO VLA Sky Survey}
\acro{OPAL}{Online Proposal Applications \& Links}
\acro{pc}{parsec\acroextra{: 1 pc $\simeq 3.09 \times 10^{16}$ m}}
\acro{PN}{Planetary Nebula}
\acro{PWN}{pulsar wind nebula}
\acro{PNe}{Planetary Nebulae}
\acro{POSSUM}{Polarisation Sky Survey of the Universe's Magnetism}
\acro{RM}{rotation measure}
\acro{QSO}{Quasi-Stellar Objects}
\acro{RFI}{Radio-Frequency Interference}
\acro{RMS}{Root Mean Squared}
\acro{SCEM}{School of Computing, Engineering and Mathematics}
\acro{SGR}{Soft Gamma Repeater}
\acro{SED}{Spectral Energy Distribution}
\acro{SFG}{Star Forming Galaxies}
\acro{SI}[$\alpha$]{Spectral Index\acroextra{, $S \propto \nu^\alpha$}}
\acro{SKA}{Square Kilometre Array}
\acro{SMB}{Super Massive Black Hole}
\acro{SMC}{Small Magellanic Cloud}
\acro{SN}{Supernova}
\acro{SNRs}{supernova remnants}
\acro{SNR}{supernova remnant}
\acro{SNe}{supernovae}
\acro{SUMSS}{Sydney University Molonglo Sky Survey}
\acro{WISE}{Wide-Field Infrared Survey Explorer}
\acro{VLA}{Very Large Array} 
\acro{VLBI}{Very Long Baseline Interferometry} 
\acro{MYSO}{Massive Young Stellar Object}
\acro{HFPs}{High Frequency Peakers}
\acro{ESP}{Early Science Project}
\acro{YSOs}{Young Stellar Objects}
\acro{IRAC}{Infrared Array Camera}
\acro{BL-Lac}{BL~Lacertae}
\acro{FSRQs}{Flat Spectrum Radio Quasars}
\acro{IR}{Infrared}
\acro{MW}{Milky Way}
\acro{MC}{Magellanic Cloud}
\acro{CARTA}{Cube Analysis and Rendering Tool for Astronomy}
\acro{GLEAM}{GaLactic and Extragalactic All-Sky MWA}
\end{acronym}

\def\papertype{\ \hfill\ Editorial}

\def\udc{52}
\def\snr{Goat's Eye}

\begin{document}
\parindent=.5cm
\baselineskip=3.8truemm
\columnsep=.5truecm
\newenvironment{lefteqnarray}{\arraycolsep=0pt\begin{eqnarray}}
{\end{eqnarray}\protect\aftergroup\ignorespaces}
\newenvironment{lefteqnarray*}{\arraycolsep=0pt\begin{eqnarray*}}
{\end{eqnarray*}\protect\aftergroup\ignorespaces}
\newenvironment{leftsubeqnarray}{\arraycolsep=0pt\begin{subeqnarray}}
{\end{subeqnarray}\protect\aftergroup\ignorespaces}
%


\markboth{\eightrm Analysis of LMC SNR N206} 
{\eightrm M. Ghavam {\lowercase{\eightit{et al.}}}}

\begin{strip}

{\ }

\vskip-1cm

\publ

\type

{\ }


\title{Multi-frequency radio-continuum study of the LMC SNR N206 (Goat's Eye) and its ``zig-zag'' PWN}


\authors{
M. Ghavam$^1$,
Z. J. Smeaton$^1$,
M. D. Filipovi\'c$^1$,
R. Z. E. Alsaberi$^{2,1}$,
C. Bordiu$^3$,
}
\authors{
W. D. Cotton$^{4, 5}$,
E. J. Crawford$^{1}$,
A. M. Hopkins$^6$, 
R. Kothes$^7$,
S. Lazarevi\'c$^{1}$,
}
\authors{
D. Leahy$^8$,
N. Rajabpour$^1$,
S. Ranasinghe$^8$,
G. P. Rowell$^9$,
H. Sano$^{2}$,
M. Sasaki$^{10}$, 
}
\authors{
D. Shobhana$^{1}$,
K. Tsuge$^{2,11}$,
D. Uro\v sevi\'c$^{12}$ and
N. F. H. Tothill$^{1}$
}

\vskip3mm


\address{$^1$Western Sydney University, Locked Bag 1797, Penrith South DC, NSW 2751, Australia}
\address{$^2$Faculty of Engineering, Gifu University, 1-1 Yanagido, Gifu 501-1193, Japan}
\address{$^3$INAF-Osservatorio Astrofisico di Catania, Via Santa Sofía 78, I-95123 Catania, Italy}
\address{$^4$National Radio Astronomy Observatory, 520 Edgemont Road, Charlottesville, VA 22903, USA}
\address{$^5$South African Radio Astronomy Observatory Liesbeek House, River Park, Gloucester Road Cape Town, 7700, South Africa}
\address{$^6$School of Mathematical and Physical Sciences, 12 Wally’s Walk, Macquarie University, NSW 2109, Australia}
\address{$^7$Dominion Radio Astrophysical Observatory, Herzberg Astronomy \& Astrophysics, National Research Council Canada, P.O. Box 248, Penticton}
\address{$^8$Department of Physics and Astronomy, University of Calgary, Calgary, Alberta, T2N IN4, Canada}
\address{$^9$School of Physics, Chemistry and Earth Sciences, The University of Adelaide, Adelaide, 5005, Australia}
\address{$^{10}$Dr Karl Remeis Observatory, Erlangen Centre for Astroparticle Physics, Friedrich-Alexander-Universit\"{a}t Erlangen-N\"{u}rnberg, Sternwartstra{\ss}e 7, 96049 Bamberg, Germany}
\address{$^{11}$Institute for Advanced Study, Gifu University, 1-1 Yanagido, Gifu 501-1193, Japan}
\address{$^{12}$Department of Astronomy, Faculty of Mathematics, University of Belgrade, Studentski trg 16, 11000 Belgrade, Serbia}


\Email{19594271@student.westernsydney.edu.au}


\dates{XXX}{XXX}


\abstract{We present new radio-continuum observations of the \ac{LMC} \ac{SNR} N206, which we give the name ``\snr''. \snr\ contains an interior radio structure that is likely a \ac{PWN}, which we analyse in further detail. We use new radio observations from the \ac{ATCA} telescope, as well as several archival radio observations, to calculate spectral indices, and find a steep spectral index for the whole \ac{SNR} ($\alpha\,=\,-0.60\pm$0.02), and a flatter spectral index for the \ac{PWN} ($\alpha\,=\,-0.16\pm$0.03). We also measure the polarisation and magnetic field properties of the \ac{PWN}. Previously reported as a linear structure, the new observations show an unusual ``zig-zag''-like structure, visible in radio-continuum total intensity, linear polarisation, and magnetic field orientations. The origin of this zig-zag structure is unclear, but we propose some origin scenarios that will require further observations to differentiate between.}



\end{strip}


\acresetall

\section{Introduction}
\label{s:intro}

The \ac{LMC} is an irregular dwarf galaxy at a distance of 50~kpc from the \ac{MW}\citep{Pietrzynski2019}. The proximity and the low Galactic foreground absorption make the \ac{LMC} an ideal laboratory for multi-frequency studies of \ac{SNR} populations in great detail. \ac{SN} explosions are important drivers of stellar and galaxy evolution, and \acp{SNR} are able to trace the interaction between the \ac{SN} ejected material and the surrounding \ac{ISM}, revealing the relationship between \ac{SNR} and galaxy evolution. \acp{SNR} in the \ac{LMC} span a wide range of evolutionary phases, from young ejecta-dominated remnants, through middle-aged adiabatic remnants, to old remnants in the radiative phase.

\begin{figure*}[ht!]    
\centering
\includegraphics[width=\textwidth]{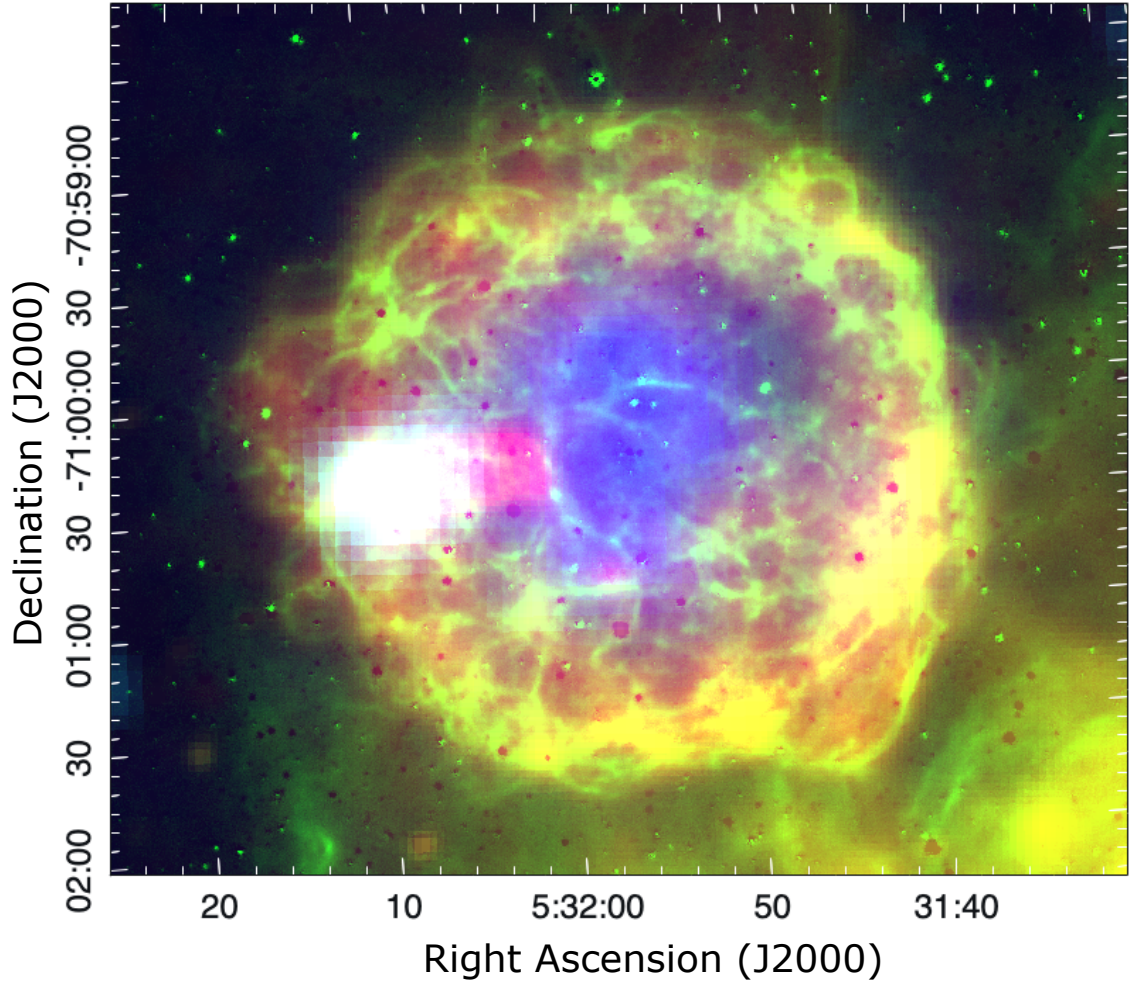}
\label{fig:RGBYM}
\caption{Multi-frequency composite colour image of \snr. Red is radio-continuum MeerKAT data (1.3\,GHz), green is optical \ac{DeMCELS} \SII\ data, blue is soft X-ray {\it XMM-Newton} data (0.2$-$1.0\,keV), yellow is optical MCELS \OIII\ data, and cyan is hard X-ray {\it XMM-Newton} data (2.0$-$4.5\,keV). All images are linearly scaled.}
\end{figure*}

Several surveys of the \ac{LMC} \ac{SNR} population have been conducted at multiple frequencies to study their overall physical properties~\citep{1996ASPC..112...91F, Filipovic1998a, Payne2007, Payne2008, 2015PKAS...30..149B, Maggi2016, Bozzetto2017, 2017ApJ...837...36L, 2021MNRAS.507.2885F, Yew2021, Bozzetto2023,2021MNRAS.507.2885F, Zangrandi2024}. In particular, radio observations are particularly useful for studying individual \acp{SNR}, advancing our understanding of the underlying physics. 

To further analyse the \ac{SNR} population of the \ac{LMC}, we present the study of the known \ac{SNR} B0532$-$71.0~\citep{mathewson1973supernova}. This \ac{SNR} is also known as \ac{SNR} N206~\citep{2002AJ....124.2135K} and is located in the high-mass star-forming complex with the same name ~\citep[also known as Henize\,206, LHA 120-N\,206, and DEM\,L221][]{Henize1956}, on the southern outskirts of the \ac{LMC}. The region surrounds the star-forming cluster NGC\,2018, also known as LHA 120-N\,206A, and contains the \HII\ region LH$\alpha$~120-N206, with the \ac{SNR} N206 located on the north-eastern edge~\citep{2004ApJS..154..275G}. N206 has been examined across multiple wavelengths by various researchers~\citep{mathewson1973supernova,Lasker1977, Milne1980,Mills1984,Williams1999,2002AJ....124.2135K,williams2005supernova,2012A&A...547A..19K}.

The \ac{SNR} N206 was initially discovered at radio frequencies~\citep{mathewson1973supernova}, and has been analysed in several subsequent radio surveys, which show a bright source with a relatively flat spectrum~\citep{Milne1980, Mills1984}; the most recent value showing  $\alpha\,=\,-0.20\pm0.07$~\citep{klinger2002peculiar}. The \ac{SNR} has also been analysed in optical, which shows a circular edge-brightened filamentary shell~\citep{Lasker1977, Long1981, williams2005supernova}, and at X-ray frequencies which shows thermal X-ray emission filling the centre of the remnant~\citep{Williams1999, williams2005supernova, 2012A&A...547A..19K}. This results in N206 being classified as a \ac{MM} \ac{SNR}~\citep{klinger2002peculiar, williams2005supernova} using the criteria of \citet{rho1998mixed}, and \citet{williams2005supernova} gives it an age of $\sim$25,000 years old, while \citet{klinger2002peculiar} estimates an older age of $\sim$29,000 years and \citet{2017ApJ...837...36L} gives an age of 13,000 years.

\ac{ATCA} radio observations of \citet{klinger2002peculiar} discovered a previously unseen linear radio feature on the east side of the remnant~\citep{klinger2002peculiar}. This feature was later detected in X-ray data from {\it Chandra} and {\it XMM-Newton}, showing associated X-ray emission with a small X-ray knot detected at the outer tip~\citep{williams2005supernova}. This feature was classified as a \ac{PWN}~\citep{klinger2002peculiar, williams2005supernova}, but the associated pulsar was not detected in following dedicated Parkes observations~\citep{williams2005supernova}.

The presence of this \ac{PWN} classifies N206 as a composite \ac{SNR}, a relatively rare class of \acp{SNR} within the \acp{MC} with only a handful of such objects known. We know of three other confirmed composite \acp{SNR} within the \ac{LMC}, 0453$-$6829~\citep{2012Harbel, 2012ApJ...756...17M}, 0540--69~\citep{2014ApJ...780...50B}, and 30\,Dor\,B~\citep{2000ApJ...540..808L}, as well as N49~\citep{2024PASA...41...89G}, N59B~\citep{2012SerAJ.184...69B}, and DEM~L316B~\citep{2005ApJ...635.1077W} which are possible composite \acp{SNR}. We know of only two within the \ac{SMC}, DEM~S5~\citep{alsaberi2019discovery} and IKT~16~\citep{2011A&A...530A.132O, 2015A&A...584A..41M, 2021MNRAS.507L...1M}.

The modern generation of radio telescopes, such as MeerKAT and ASKAP, have allowed the discovery of several new \acp{SNR} and \ac{SNR} candidates, both in our Galaxy~\citep{Kothes2017, Filipovic2023, BurgerSchiedlin2024, Lazarevic2024, Perun, Smeaton2024, 2025arXiv250504041F} and in the \acp{MC}~\citep{Bozzetto2023, 2024MNRAS.529.2443C, Yew2021}, as well as analyses which allow us to better understand the physical properties of already known \acp{SNR}~\citep[]{Diprotodon,2025arXiv250615067S}. The improved sensitivity and angular resolution also allow better analysis of a number of different astrophysical objects~\citep[]{2022MNRAS.516.1865V, 2023MNRAS.523.1933V, 2024A&A...690A..53B, 2024PASA...41...32L, 2025arXiv250606768A, Bradley2025, 2025ApJ...984L..52F}.

In this paper, we add to these analyses by presenting radio-continuum observations of the \ac{SNR} N206, which we now give the nickname ``\snr''\footnote{The name ``\snr'' comes from the radio-continuum morphology which shows a circular shell with the bright linear \ac{PWN}, which resembles a spherical eye with a large and unique horizontal/rectangular pupil as seen in goats.}. The newly presented radio data come from the \ac{ATCA}, MeerKAT, and \ac{ASKAP} telescopes at frequencies ranging from 944\,MHz to 9\,GHz. The new high-resolution radio data shows more detail in the \ac{PWN}, revealing a previously unseen ``zig-zag'' structure which we investigate further. We also present archival radio-continuum measurements from \ac{MWA}, \ac{MOST}, and combined \ac{ATCA} and Parkes mosaics, as well as previous optical and X-ray data to provide an overall multi-frequency view. We specifically analyse the properties of the linear \ac{PWN} structure. In Section~\ref{s:Data, Observations and Processing}, we present the data used and observation details. In Section~\ref{s:ResultsDiscussion}, we present the results and discussion, while in Section~\ref{sec:conclusion}, we present our conclusions.

\begin{figure*}[ht!]    
\centering
\includegraphics[width=0.97\textwidth]{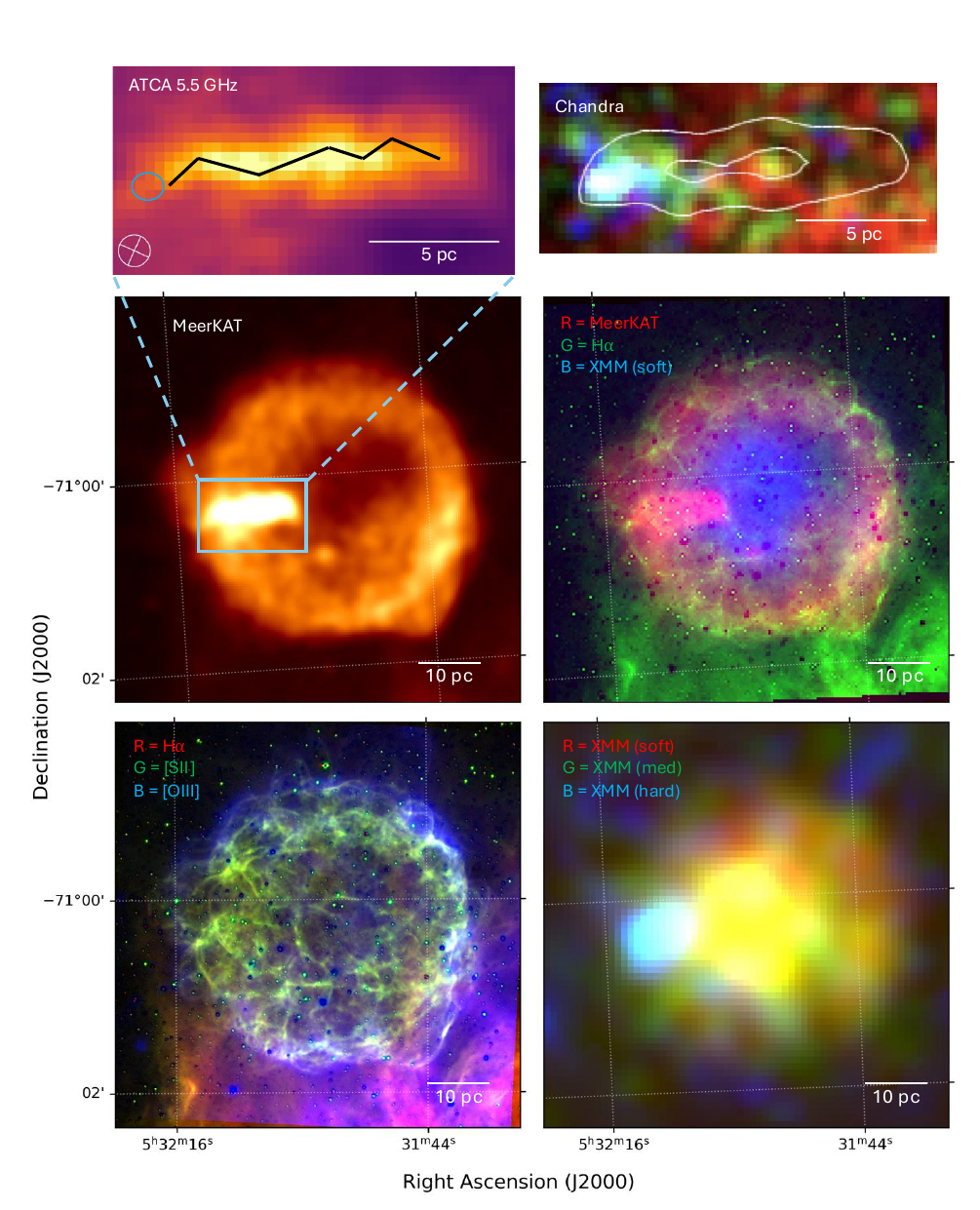}
\caption{4-panel image of \snr. All images have linear scaling and display a relevant scale bar in the bottom right corner. {\bf Top left:} MeerKAT radio-continuum image at 1.3\,GHz. The left inset shows a zoomed-in \ac{ATCA} 5.5\,GHz view of the \ac{PWN}. The described ``zig-zag'' structure is annotated with the black line and the position of the X-ray point source possible pulsar is shown in the blue circle. The beam size of 5\arcsec$\times$5\arcsec\ is shown in the bottom left corner. The top right image shows the same \ac{PWN} view using an X-ray RGB made with {\it Chandra} data. Red is soft band (0.5$-$1.2\,keV), green is medium band (1.2$-$2.0\,keV), and blue is hard band (2.0$-$7.0\,keV). Contours are from the left-hand \ac{ATCA} image at levels of 0.4 and 0.8 mJy beam$^{-1}$. {\bf Top right:} Multi-frequency RGB where red is MeerKAT radio at 1.3\,GHz, green is \ac{DeMCELS} H$\alpha$, and blue is {\it XMM--Newton} soft X-ray (0.2--1.0\,keV). {\bf Bottom left:} Optical RGB where red is \ac{DeMCELS} H$\alpha$, green is \ac{DeMCELS} \SII\, and blue is MCELS \OIII. {\bf Bottom right:} {\it XMM--Newton} X-ray RGB where red is soft X-ray (0.2--1.0\,keV), green is medium X-ray (1.0--2.0\,kev), and blue is hard X-ray (2.0--4.5\,keV).}
\label{fig:4panels}
\end{figure*}

\section{Observations and Data Reduction}
\label{s:Data, Observations and Processing}

\subsection{Radio}
\label{subsec:radio}

\subsubsection{ATCA data and processing}
\label{subsubsec:ATCA}

Here, we present details from our new \ac{ATCA} observations at 5.5 and 9~GHz, which were acquired on 4$^\mathrm{th}$~December~2019 as part of project CX403. The observations were carried out using a 1.5C array configuration, with one hour of integration over a minimum 12-hour period. The observations used the \ac{CABB} (with 2048\,MHz bandwidth) at wavelengths of 3 and 6\,cm ($\nu$=4.5--6.5 and 8--10\,GHz; centred at 5500 and 9000\,MHz). We employed the standard southern-sky calibrator PKS~B1934--638 for the primary (flux density) calibration, and source QSO~J0047--7530 for the secondary (phase) calibration. Details of these calibrators can be found in the \ac{ATCA} calibrator database\footnote{\url{https://www.narrabri.atnf.csiro.au/calibrators/calibrator\_database.html}}. The data contain both total power Stokes~$I$ observations and linear polarisation data, which are used in Section~\ref{subsec:polarisation}. 

The data were reduced and analysed using {\sc MIRIAD}~\citep{1995ASPC...77..433S} and {\sc KARMA} \citep{1995Gooch} software packages. We produced and deconvolved images by setting the robust weighting parameter to 0.5 (for all 5.5 and 9\,GHz images) and deconvolving via a primary beam correction. Both final Stokes~$I$ images achieve a resolution of 5\arcsec$\times$5\arcsec\ and a local \ac{RMS} noise level of $\sim$14\,$\mu$Jy\,beam$^{-1}$ for 5.5\,GHz and $\sim$10$\,\mu$Jy\,beam$^{-1}$ for 9\,GHz.

We also used previous \ac{ATCA} data in the form of archival mosaic images generated from combining \ac{ATCA} and Parkes radio-continuum data for some flux density measurements (see Table~\ref{tabflux}). The properties and generation of these mosaics are described in further detail in \citet{Filipovic1995PaperIV, Filipovic1998PaperVII, 2005AJ....129..790D, 2007MNRAS.382..543H, 2021MNRAS.507.2885F}.

\subsubsection{Other radio data}
\label{subsubsec:MeerKAT}

The MeerKAT data is obtained from the most recent \ac{LMC} MeerKAT survey (project code SSV-20180505-FC-02). We make use of the broadband image from this dataset, centred at 1295\,MHz over the bandwidth 856$-$1712\,MHz (see Fig.~\ref{fig:4panels}, top left panel). Further image details, including calibration, data reduction, and the final survey information, will be presented in Cotton et al. (in prep) and Rajabpour et al. (in prep). 

The \ac{ASKAP} radio-continuum data primarily comes from the \ac{LMC} survey conducted as part of the \ac{ASKAP} commissioning and early science (ACES, project code AS033)~\citep{2021MNRAS.506.3540P, Bozzetto2023}. This observation was conducted using the entire 36-antenna array at 888\,MHz with a 288\,MHz bandwidth. The data has a restoring beam of 13\farcs9$\times$12\farcs1 (position angle\ PA=--84.4\D) and was reduced using the standard ASKAPSoft pipeline including multi-scale cleaning, self-calibration, and multi-frequency synthesis imaging~\citep{Guzman2019}. We also use a 944\,MHz flux density measurement from Smeaton et al. (in prep.), which was measured from the \ac{ASKAP} \ac{EMU} survey data of the \ac{LMC}~\citep{2025PASA...42...71H}.

We use radio-continuum data from the \ac{MWA}~\citep{Tingay2013} \ac{GLEAM} survey~\citep{Wayth2015, HurleyWalker2017} of the \ac{LMC}. We use the images at 118, 155, and 200\,MHz as described in \citet{2018MNRAS.480.2743F} to measure the \ac{SNR} flux densities.

We also use radio-continuum images from the \ac{SUMSS} conducted with the \ac{MOST} telescope to measure the flux density at 843\,MHz. Details of this data and survey can be found in \citet{2003MNRAS.342.1117M}.

\begin{figure*}[ht!]
\centering
\includegraphics[width=\linewidth]{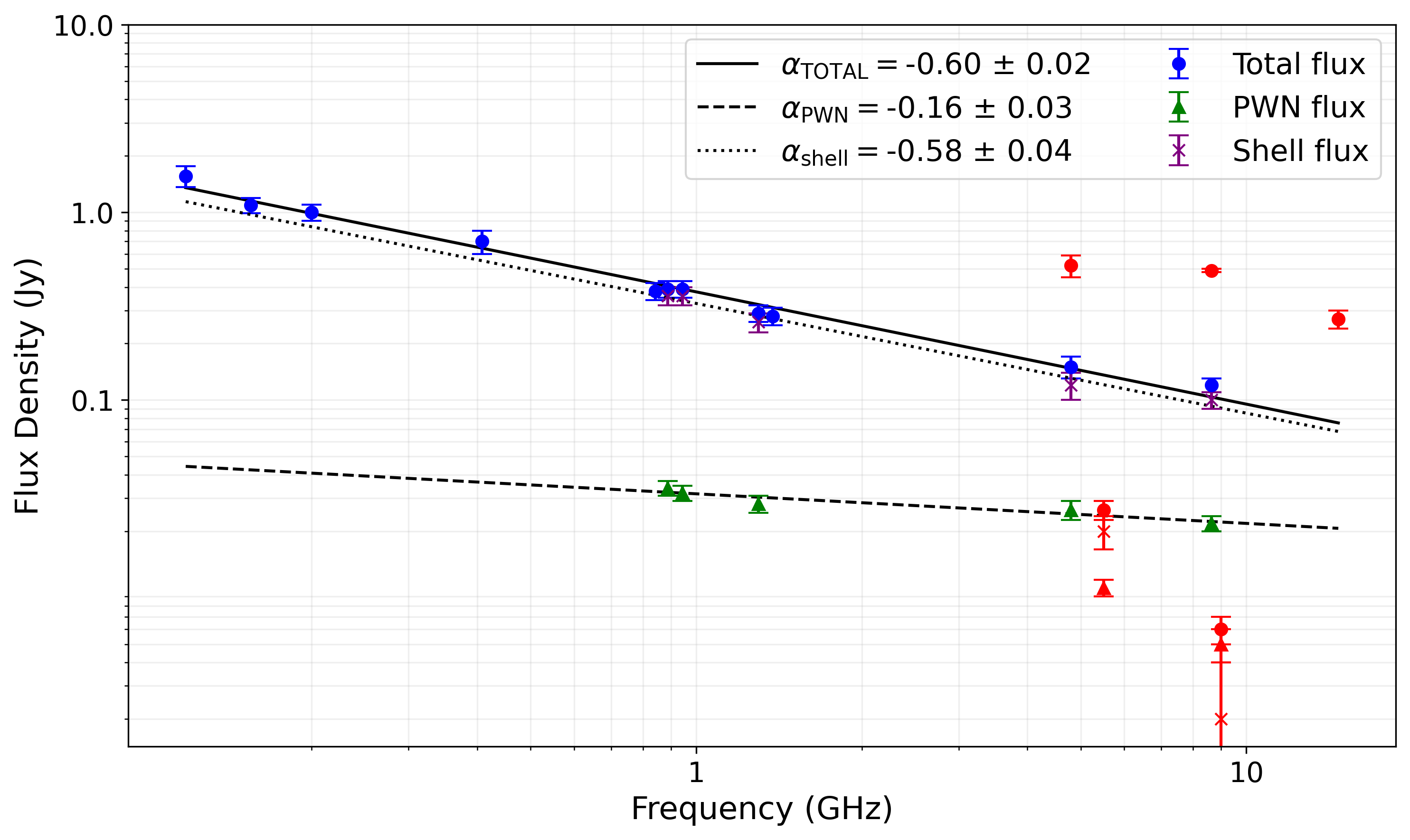}
\caption{Radio-continuum spectrum of \snr\ using radio-continuum data from (Table~\ref{tabflux}). The black solid line gives the total spectral index, the black dashed line gives the PWN spectral index, and the black dotted line gives the shell spectral index, as shown in the legend. The points in red are the non-fitted data in Table~\ref{tabflux}. }
\label{fig:SIgraph}
\end{figure*}

Archival \HI\ data~\citep{2003ApJS..148..473K} were obtained using the \ac{ATCA} and Parkes 64-m telescope. The angular resolution of the data is 60\arcsec, corresponding to a spatial resolution of $\sim$15\,pc at the distance of the \ac{LMC}~\citep{Pietrzynski2019}. The typical noise level is $\sim$2.4\,K at a velocity resolution of 1.689\,km\,s$^{-1}$.

\subsection{Optical}
\label{subsec:optical}

The optical data shown is from the \ac{MCELS} and \ac{DeMCELS} optical surveys~\citep{1999IAUS..190...28S, 2024ApJ...974...70P}, which imaged the \ac{LMC} using the UM/CTIO Curtis Schmidt telescope. The \ac{DeMCELS} survey achieves a higher resolution, but does not include images using the \OIII\ filter. Therefore, the images used are the H$\alpha$ and \SII\ images from the \ac{DeMCELS} survey, and the \OIII\ image from the \ac{MCELS} survey (see Fig.~\ref{fig:4panels}, bottom left panel).

\subsection{X-Ray}
\label{subsec:X-ray}

The X-ray data shown are from the {\it XMM-Newton} telescope survey of the \acp{MC}~\citep{Haberl2014, Maggi2016, Maggi2019}, and from the {\it Chandra} telescope. The {\it XMM} survey used a bandwidth 0.2$-$10.0\,keV and achieved a sensitivity of $F_\mathrm{X}$ (0.3$-$8\,keV) $\approx 10^{-14}$ erg\,cm$^{-2}$\,s$^{-1}$. We show all three X-ray bands, soft, medium, and hard (see Fig.~\ref{fig:4panels}, bottom right panel). The {\it Chandra} data achieves a higher resolution but a lower sensitivity than {\it XMM-Newton}, and is therefore less sensitive to the more diffuse shell structure. While the shell is visible in the soft {\it Chandra} band, this diffuse emission is better seen in the {\it XMM-Newton} data, and the {\it Chandra} data is primarily used to analyse the \ac{PWN} with higher resolution and specifically show the X-ray point source (see Fig.~\ref{fig:4panels}, top right inset). We use all three {\it Chandra} bands, soft (0.5$-$1.2\,keV), medium (1.2$-$2.0\,keV), and hard (2.0$-$7.0\,keV), and the images were generated by merging together two separate {\it Chandra} observations at each band.

\section{Results and Discussions}
\label{s:ResultsDiscussion}
\subsection{Morphology}
\label{subsec:radio morphology}

\begin{figure*}[ht!]
\centering
\includegraphics[width=\textwidth]{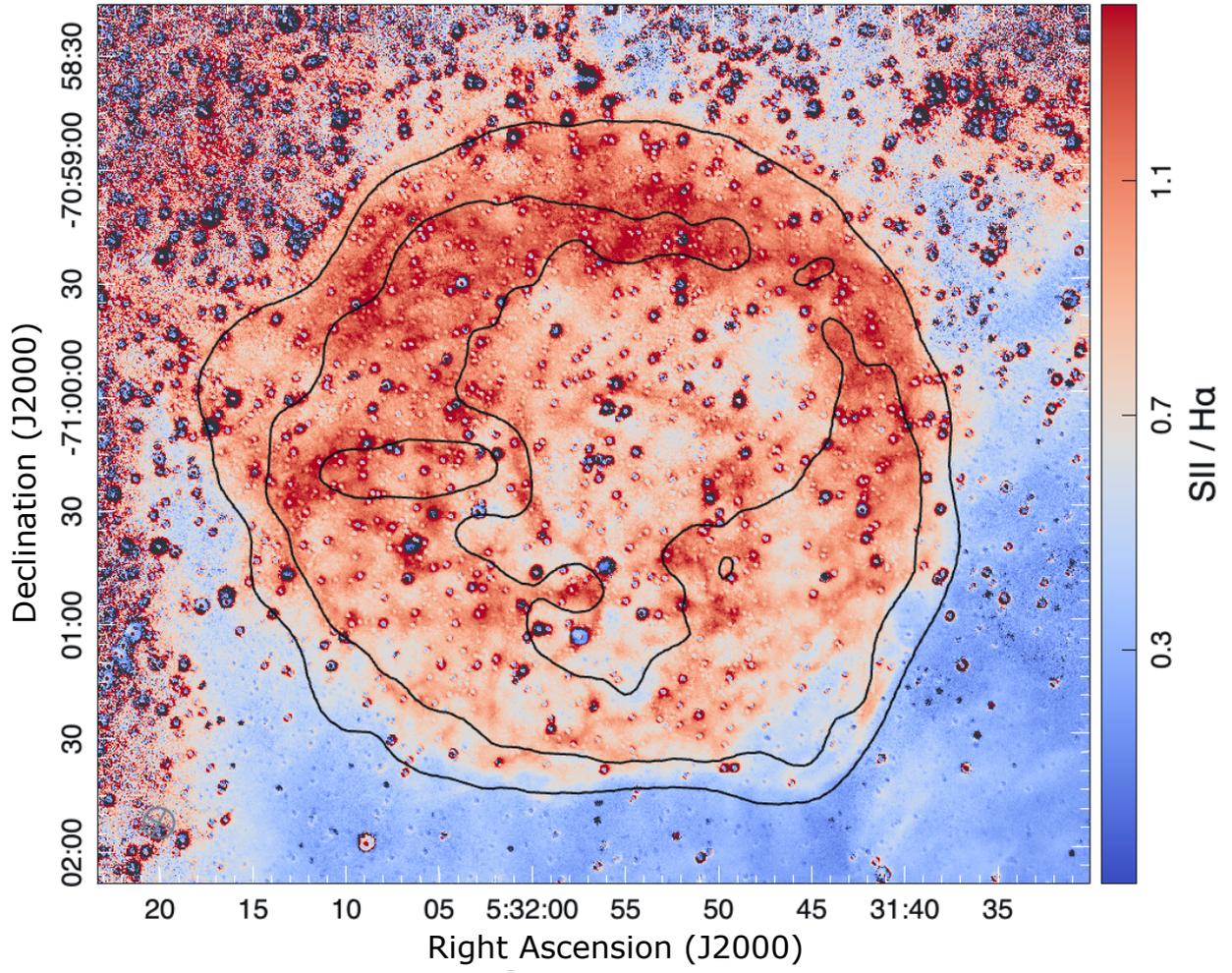}
\caption{ \SII/\Halpha\ ratio map of \snr\ from optical \ac{DeMCELS} data. The contours are from the MeerKAT radio-continuum image at levels of 0.3, 0.8, and 2.0 mJy beam$^{-1}$.}
\label{fig:S2H}
\end{figure*}


\subsubsection{\snr\ Shell}
\label{subsubsec:shell}

The radio-continuum images show a circular, edge-brightened, shell-type morphology (see Fig.~\ref{fig:4panels}, top left panel) with the bright linear radio feature identified by \citet{klinger2002peculiar} in the eastern side. There are also two areas, on the east and south-west of the \ac{SNR}, which deviate slightly from this circular shape. These areas are potentially small blowout structures, sometimes referred to as ``ears''~\citep{2021MNRAS.502..176C}, as seen in some other \acp{SNR} (e.g. G1.9+0.3 \citep{Luken2020}, SNR~S147~\citep{Xiao2008}, SNR~J0455$-$6838~\citep{Crawford2008}). The edge-brightened shell is fairly symmetrical, with the exception of the brighter south-west rim. 

From the MeerKAT data, we measure the \ac{SNR} by eye, and find a geometric centre to be at RA(J2000) = 05:31:56.6, Dec(J2000) = $-$71:00:17, with an angular size of 3\farcm0$\times$3\farcm1, corresponding to a physical size of 44$\times$45\,pc at the \ac{LMC} distance of 50\,kpc~\citep{Pietrzynski2019}. This is slightly smaller than that reported in \citet{klinger2002peculiar} from their \ac{ATCA} data (44\,pc$\times$47\,pc), however this is likely due to fitting the region by eye as exact region parameters were not previously reported. This physical size is slightly larger than the average diameter for the \ac{LMC} population~\citep[mean diameter of 41\,pc for the \ac{LMC};][]{Bozzetto2017}, but still well within the population distribution for an \ac{SNR}.

\snr\ also has an observed optical shell, with an inner filamentary structure seen in \Halpha\ and \SII, and an outer \OIII\ rim. We detect bright interior thermal X-ray emission in the soft X-ray band, but do not detect the outer shell. The spatial structure of the \SII/\Halpha\ shell closely matches the radio-continuum shell in the MeerKAT image, with the \OIII\ extending slightly beyond this emission. This \SII\ emission extends across the entire \ac{SNR} and matches the radio-continuum extent, indicating that the shockwave is composed of predominantly radiative shocks. The two potential blowout regions seen in the radio-continuum images are also seen in \Halpha\ and \SII, indicating that they are components of the expanding shell structure. The \Halpha\ and \SII\ emission shows a filamentary web-like structure over the \ac{SNR}, which may be indicative of expansion into a porous \ac{ISM}~\citep{Dimaratos2015}. 

Such filamentary optical structures are not unusual in \acp{SNR}, such as 0450$-$7050~\citep{2025arXiv250615067S}. The point that makes Goat's Eye's morphology so unusual is the \OIII\ emission being located outside of the \SII/\Halpha\ shell, as it is typically located closer to the interior where the temperature is often higher, or at areas of faster shock velocities where significant interaction may be occurring. The fact that we see an \OIII\ shell around the entire \ac{SNR} leading the slower, radiative shell seen in \SII\ and \Halpha\ indicates there may be a complex velocity structure in the shell. If the \ac{SN} occurred in a cavity which was cleared by the progenitor wind, then this may help to explain the observed properties. The \OIII\ may indicate the point where the shock is just reaching the end of the cavity and is hitting the cavity walls with sufficient velocity to emit \OIII\ emission. This may also help to explain the extremely filamentary structure visible in \SII\ and \Halpha, since the pre-excavated cavity could have created a complex, porous density structure resulting in the filamentary structure as the shock waves preferentially expand into the lowest density paths. 

There is also a thick shell observed in radio, approximately 30\arcsec\ inward for the thickest shell. There are multiple possible reasons for this thicker shell. Potentially, as the expanding shock wave is encountering the cavity wall, it begins to bounce back and slow down, travelling back toward the \ac{SNR} interior. However, with lower shock velocities as demonstrated by the \SII/\Halpha\ correlation. This material would then be compressed on both sides, and thus could be stacked up and form a thick inner shell. Additionally, as it expands inwards, it will encounter the \ac{SN} ejecta, which may result in enhanced mixing and cause a turbulent zone inside the outer shell.

In Fig.~\ref{fig:S2H} we display the \SII/\Halpha\ ratio map generated from the optical \ac{DeMCELS} data. We measure an average of $\sim$1.1$\pm$0.4 over the entire shell, with higher values of $\sim$1.2$-$1.3 in the northern half. This suggests that the northern shock may have slowed down and is becoming radiative, or has enhanced cooling in this region \citep[see also][]{2015MNRAS.454..991R,2014Ap&SS.351..207B}. The southern shock, oriented toward the \HII\ region, shows radio edge brightening, indicating potential interaction.

We also checked the available \ac{IR} {\it Herschel} and {\it Spitzer} images of the \ac{LMC}~\citep{Meixner2013}, and do not see \snr's shell visible. This is expected from the results of \citet{2015ApJ...799...50L}, who found no significant dust heating or interaction in their \ac{IR} analysis. Therefore, if there is interaction between the \ac{SNR} shell and the nearby \HII\ region, it may be just beginning, or if the shocks are particularly slow, then they may not be energetic enough to cause significant dust heating. Alternatively, the \ac{SNR} and the \HII\ region may be separated by an offset along the line of sight, which would explain the lack of observed interaction.


\subsubsection{Shell Symmetry}
\label{subsubsec:symmetry}

To quantify the \snr's shell symmetry, we measure the multipole values from the MeerKAT radio-continuum images and compare with the results of \citet{Multipole2019} and \citet{2025PASP..137f4502L}. Using the method of \citet{Multipole2019} we measure values of P$_2$/P$_0$ = 280.23$\pm$0.51$\times$10$^{-6}$ and P$_3$/P$_0$ = 16.596$\pm$0.063$\times$10$^{-6}$, which place \snr\ in the centre of the \ac{SNR} population studied in \citet{Multipole2019}. Therefore, this comparison is not definitive in classifying \snr's morphology. 

Following the newer method of \citet{2025PASP..137f4502L}, we calculate the radial and angular dipoles, quadrupoles, and octopoles separately. We find that \snr's radial component has a low dipole moment and is dominated by the quadrupole moment, and the angular component is dominated by the dipole moment. This matches the radio morphology displayed in Fig.~\ref{fig:4panels}, where the main deviation in radial asymmetry is caused by the two small blowout structures seen in the south-west and north-east. These structures are resulting in two directions of asymmetry, and thus a larger quadrupole moment. The higher angular dipole moment is likely being caused by the bright \ac{PWN} in the east, and thus this indicates that this is the dominant cause of angular asymmetry in \snr. \citet{2025PASP..137f4502L} primarily used X-ray morphologies, and we note that the radio multipoles are significantly different from the X-ray multipoles reported for \snr, indicating that there are significant differences in the radio and X-ray morphology, as observed in our images (see Fig.~\ref{fig:4panels}).

This low radial dipole moment, and the observed radio morphology, show a remarkably circular structure, similar to some of the most circular \acp{SNR}, such as the Galactic \ac{SNR} Teleios~\citep{2025arXiv250504041F} and the circumgalactic \ac{SNR} J0624$-$6948~\citep{2022MNRAS.512..265F,2025A&A...693L..15S}. For such an evolved~\citep[$\sim$25,000$-$29,000 years]{klinger2002peculiar, williams2005supernova} \ac{CC} \ac{SNR}, this remarkable circularity is unusual. \snr\ undoubtedly originates from a \ac{CC} \ac{SN} due to the presence of the \ac{PWN}. \ac{CC} \acp{SN} are expected to have asymmetrical explosions~\citep{2009ApJ...706L.106L, 2011ApJ...732..114L}, and thus asymmetrical remnants, as opposed to their more symmetrical Type~Ia counterparts. However, this possible correlation is debated in the literature~\citep{2019arXiv190911803R, 2025PASP..137f4502L}. Additionally, \acp{SNR} are expected to remain more circular when they expand into more rarefied and homogeneous environments, for example, in the cases of Teleios and J0624$-$6948. The nearby \HII\ star-forming region to the south-west makes this scenario less likely for \snr, however it is possible that there is an offset between \snr\ and the \HII\ region which would account for the lack of interaction signatures. 
If \snr\ is in close physical proximity to the neighbouring region, then it is unusual that \snr\ was able to retain such a circular shape throughout its 
evolution, and this would require detailed modelling that accounts for the complex environment and explosion subtype.

\subsubsection{\snr's PWN}
\label{subsubsec:PWNFeature}

The finer structure of the \ac{PWN} feature is shown in the higher-resolution \ac{ATCA} data (see Fig.~\ref{fig:4panels}, inset), with a measured length of 50\arcsec, corresponding to a physical length of $\sim$12\,pc for the \ac{LMC}. Originally described as a linear feature in \citet{klinger2002peculiar}, the new \ac{ATCA} observations show a more ``zig-zag''-like structure. This proposed structure is highlighted in the inset image by the annotated black line.
There is also a bright, non-thermal, X-ray point source located at the east side of the \ac{PWN}, reported in \citet{williams2005supernova}. This source is visible in the {\it Chandra} X-ray images (see Fig.~\ref{fig:4panels}, top right inset), and is clearly located within the bounds of the \ac{PWN} as shown by the radio contour levels. There is no corresponding radio point source visible.

This structure was suggested to be a run-away, or bow-shock, \ac{PWN}~\citep{klinger2002peculiar}, wherein a pulsar is given a high kick velocity during the initial \ac{SN} explosion. This runaway pulsar then travels supersonically, generating a tail trailing behind which can emit strongly at X-ray and radio frequencies. This scenario was explored in more detail in \citet{klinger2002peculiar}. The morphology of \snr's \ac{PWN} supports this scenario as it shows similarities to other known bow-shock \ac{PWN}, such as DEM~S5 in the \ac{SMC}~\citep{alsaberi2019discovery}, the Galactic Lighthouse \ac{PWN}~\citep[IGR J11014-6103][]{2014IJMPS..2860172P, 2014A&A...562A.122P, 2016A&A...591A..91P}, the Galactic Potoroo \ac{PWN}~\citep{2024PASA...41...32L}, and the Galactic Mouse \ac{PWN}~\citep{1987Natur.330..455Y, 2002ApJ...579L..25C}. \snr\ appears to be particularly similar to the Galactic Mouse \ac{PWN}, which also appears to be moving away from the centre of a nearby \ac{SNR} G359.1$-$0.5. Recent results however, argue that the Mouse \ac{PWN} is not physically associated with G359.1$-$0.5 as the pulsar is moving too slowly and thus is too old to have originated from the \ac{SNR}~\citep{2009ApJ...706.1316H}. 

In the case of \snr's bow-shock \ac{PWN}, the head of the bow-shock structure would contain the high-velocity pulsar, on the eastern side travelling out of the \ac{SNR}, and the tail pointing back toward the geometric centre of the \ac{PWN}; the origin point of the pulsar. An X-ray point source is detected on the eastern side of the \ac{PWN}~\citep{williams2005supernova}, which may represent the pulsar. This is unconfirmed however, as previous timing searches did not detect the presence of a pulsar~\citep{williams2005supernova}. We do not detect any radio counterpart to this source, or detect any compact radio sources within the \ac{PWN}. It is possible that the pulsar is present, but radio quiet, or that the presence of the pulsar is hidden within the \ac{PWN} emission. \citet{klinger2002peculiar} estimated a tangential velocity of $\sim$800\,km\,s$^{-1}$, which is a reasonable velocity value for a bow-shock \ac{PWN}, and thus the \snr\ pulsar may have been born at the time of the \ac{SN} explosion and be physically associated.

\subsubsection{Distribution of {\rm \HI} clouds}

\begin{figure*}[ht!]
\centering
\includegraphics[width=\textwidth]{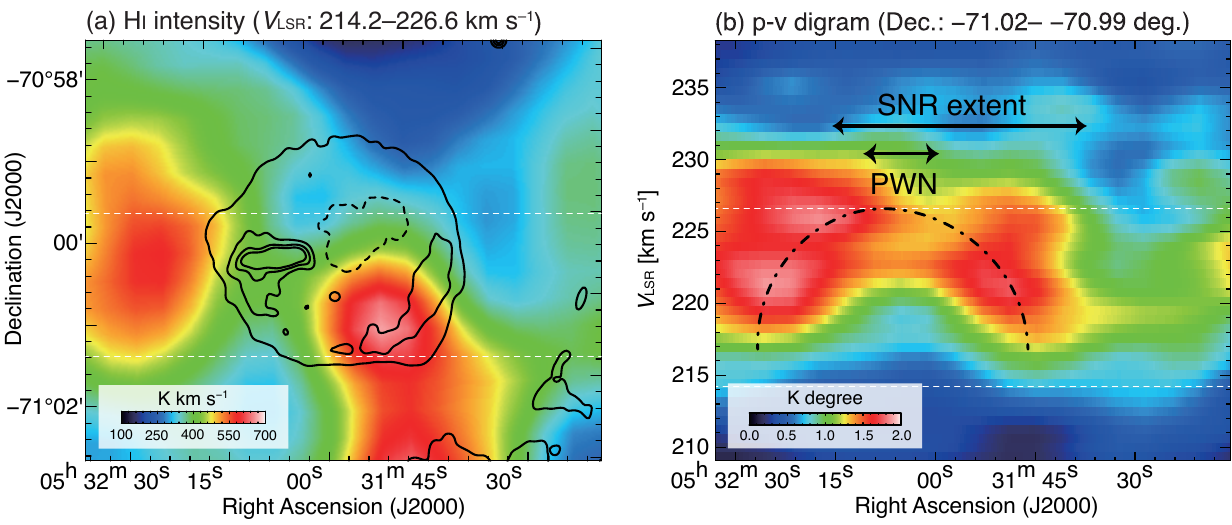}
\caption{\HI\ integrated intensity map and $p-v$ diagram of \snr\ and surrounding environment. {\bf Panel (a): }\HI\ intensity map of \snr\ and surrounding environment integrated over the $V_\mathrm{LSR}$ velocity range of 214.2--226.6\,km\,s$^{-1}$. Black contours are from MeerKAT 1.3\,GHz radio-continuum map at levels of 0.5, 1.0, 1.5, and 2.0\,mJy\,beam$^{-1}$. The dashed white horizontal lines correspond with panel (b) and show the extent of the possible \HI\ cavity. {\bf Panel (b): }\HI\ $p$--$v$ diagram, integrated over \snr's Dec location ($-$71\fdg02--\,\,--70\fdg99). The dashed white horizontal lines correspond with panel (a) showing the location of the cavity relative to \snr. The black dashed arc traces the possible \HI\ cavity, and the labelled arrows display the \ac{PWN} and \ac{SNR} physical extents.}
\label{fig:HI}
\end{figure*}

The spatial distribution of \HI\ emission toward \snr\ is presented in Fig.~\ref{fig:HI} (panel a). An \HI\ cloud is identified in the south-western part of the remnant, coinciding with a region of enhanced radio continuum emission. Another \HI\ structure is observed along the eastern shell. In contrast, there is less \HI\ emission visible in the north-western region. This lower emission corresponds with a more circular shell structure with lower radio-continuum emission. This supports the scenario that the enhanced radio-continuum emission and slight asymmetries are due to interaction with neighbouring \HI\ regions which is not present in this area.

Fig.~\ref{fig:HI} (panel b) shows the position--velocity ($p-v$) diagram of \HI\ emission toward \snr. There is a possible arc-like feature which appears to trace the peaks of \HI\ in the $p-v$ diagram. This possible feature spans a spatial diameter of 44\,pc and has a velocity width of $\sim$10\,km\,s$^{-1}$. While the extent of the radio shell does not fully coincide with this possible \HI\ structure, it appears that the geometric centre of the expanding \HI\ shell potentially aligns with the spatial position of the \ac{PWN}.

Given the low angular resolution of the \HI\ data (60\arcsec) it is difficult to determine if this is a true alignment. The angular size of the shell is $\sim$3\arcmin, approximately 3 times the \HI\ angular resolution, and the \ac{PWN} is $\sim1\arcmin$. Therefore, the large-scale \HI\ distribution is likely reliable, but the smaller-scale features related to the \ac{PWN} are less definitive as they are at the same spatial scale as the \HI\ resolution. It is possible that the larger-scale structure was formed by stellar winds from a massive progenitor. If this is the case, then it is supported  by the measured velocity width of $\sim$10\,km\,s$^{-1}$, which is reasonable for these kinds of structures~\citep{2021ApJ...923...15S, 2022ApJ...933..157S, 2024ApJ...961..162F}.

\subsection{Flux density}
\label{subsec:flux density}

\begin{table*}
\footnotesize
\caption{Flux density measurements of \snr\ at multiple radio frequencies. $^\star$ indicates that a quoted uncertainty was not given. $\dag$ indicates flux density measurements which are included in the table, but which were not used in the spectral index estimate, see text for details.}
\begin{tabular}{@{}cccccl@{}}
\hline\hline
$\nu$ & $S_{\text{Total}}$ & $S_{\text{PWN}}$ & $S_{\text{Shell}}$ & Telescope & Reference\\
(MHz) & (Jy) & (Jy) & (Jy) &  & \\
\hline
118&1.56$\pm$0.2&$-$&$-$&MWA&This work (from \citet{2018MNRAS.480.2743F} image)\\
155&1.09$\pm$0.1&$-$&$-$&MWA&This work (from \citet{2018MNRAS.480.2743F} image)\\
200&1.00$\pm$0.10&$-$&$-$&MWA&This work (from \citet{2018MNRAS.480.2743F} image)\\
408&0.7$\pm-^\star$&$-$&$-$&MOST&\citet{mathewson1973supernova}\\
843&0.38$\pm$0.04&$-$&$-$&MOST&This work (from \citet{2003MNRAS.342.1117M} image)\\
888&0.39$\pm$0.04&0.034$\pm$0.003&0.36$\pm$0.04&ASKAP&This work (from \citet{2021MNRAS.506.3540P} image)\\
944&0.39$\pm$0.04&0.032$\pm$0.003&0.36$\pm$0.04&ASKAP& Smeaton et al. (in prep)\\
1298&0.29$\pm$0.03&0.028$\pm$0.003&0.26$\pm$0.03&MeerKAT&This work (from Cotton et al. (in prep) image\\
1377&0.28$\pm$0.03&$-$&$-$&ATCA + Parkes&This work (from \citet{2007MNRAS.382..543H} image)\\
4798&0.52$\pm$0.07$\dag$&$-$&$-$&ATCA&\citet{klinger2002peculiar}\\
4800&0.15$\pm$0.02&0.026$\pm$0.003&0.12$\pm$0.02&ATCA + Parkes&This work (from \citet{2005AJ....129..790D} image)\\
5500&0.030$\pm$0.003$\dag$&0.010$\pm$0.001$\dag$&0.020$\pm$0.004&ATCA&This work\\
8638&0.49$\pm$0.12$\dag$&$-$&$-$&ATCA&\citet{klinger2002peculiar}\\
8640&0.12$\pm$0.01&0.022$\pm$0.002&0.10$\pm$0.01&ATCA + Parkes&This work (from \citet{2005AJ....129..790D} image)\\
9000&0.007$\pm$0.001$\dag$&0.005$\pm$0.001$\dag$&0.002$\pm$0.002&ATCA&This work\\
14700&0.27$\pm-^\star\dag$&$-$&$-$&Parkes&\citet{Milne1980}\\
\hline
\hline
& $\alpha_\text{Total}$ & $\alpha_\text{PWN}$ &\textbf{$\alpha_\text{Shell}$}&&\\
& $-0.60\pm0.02$ & $-0.16\pm0.03$ & \textbf{-0.58$\pm$0.04}&\\ \hline
\end{tabular}
\label{tabflux}

\end{table*}

We measure the flux density of both the entire \ac{SNR} from all available radio frequencies, and of the \ac{PWN} for data with sufficiently good resolution. We also subtracted the nearby background emission when measuring flux density, as explained in \citet{2019PASA...36...48H}. This process was done by fitting a circular region around the entire shell and an elliptical region around the \ac{PWN} using the astronomy imaging software \ac{CARTA}~\citep{CARTA_2018}. These same regions is used for all the flux density measurements shown in Table~\ref{tabflux}. The errors are taken as 10\% following a similar process to that used in \citet{2022MNRAS.512..265F, Filipovic2023, Diprotodon, Smeaton2024,Bradley2025} and \citet{2025PASA...42...69A}. We also calculate the flux density of the shell by taking it as the difference between the total and \ac{PWN} flux densities.



\subsection{Spectral Index}
\label{subsec:spectral index}

We use the flux density measurements shown in Table~\ref{tabflux} to calculate the spectral index of \snr, where the spectral index is defined as $S\propto\nu^\alpha$~\citep{2021pma..book.....F}. We calculate the spectral index separately for both the total \ac{SNR} and for the \ac{PWN} (see Fig.~\ref{fig:SIgraph}). The fitting is done using the \textsc{linregress}\footnote{\url{https://docs.scipy.org/doc/scipy/reference/generated/scipy.stats.linregress.html}} function from the \textsc{scipy} Python library~\citep{Virtanen2020}. 

Initially, all radio-continuum data points were used for the fitting, and there were some points that were significant outliers to the fit. To quantify this, we measured the reduced $\chi^2$ value as the sum of (residuals/rms)$^2$, where residuals = measured value $-$ model value. We calculate a reduced $\chi^2$ value of 60.37 for the fit. The main outliers were the \ac{ATCA} data points; the 4798 and 8638\,MHz data points from \citet{klinger2002peculiar} were much higher than the fit, and the 5500 and 9000\,MHz data measured in this research (project CX403) were much lower. We investigated our \ac{ATCA} data and found that there were significant missing short spaces, which are likely resulting in a flux density underestimation. To mitigate this, we searched the mosaic images of \citet{2005AJ....129..790D}, which combines the \ac{ATCA} interferometry data with single-dish Parkes data to fill in the missing short spacings. We found merged images at 4800 and 8640\,MHz, which are similar frequencies to the outlying points. These new points fit the line much better and thus are used instead of the \ac{ATCA} outlying points. We also exclude the 14700\,MHz data point from \citet{Milne1980}, as it is also an outlier, likely due to lower sensitivity and resolution from the older survey. After excluding these points, we achieve a reduced $\chi^2$ value of 0.88, indicating a much better fit for the data.

We also fit a separate spectral index for just the \ac{PWN}. This fit has many fewer data points, as the \ac{PWN} can only be resolved in the highest-resolution radio images (see Table~\ref{tabflux}). For this fit, we also exclude the 5500 and 9000\,MHz data points, as these are outliers to the fit and disagree with the combined \ac{ATCA} and Parkes data at similar frequencies, similar to the argument above. This is also shown by the reduced $\chi^2$ values which is 20.97 when these two points are included, and is 0.40 when they are excluded, indicating that their exclusion results in a much better fit for the data. The shell spectral index is calculated in a similar way, with the shell flux densities being calculated as the difference between the total and \ac{PWN} flux densities. Similarly, the 5500 and 9000\,MHz data points are excluded from the fit as they are outliers to the fit, with the reduced $\chi^2$ value being 71.94 when they are included, and 0.55 when they are excluded.

These results are shown in Fig~\ref{fig:SIgraph}, where the fitted points are shown in blue (for the total), green (for the \ac{PWN}), and purple (for the shell), and all the non-fitted points are shown in red. We calculate spectral index values of $-0.60\pm0.02$ for the total, $-0.16\pm0.03$ for the \ac{PWN}, and $-0.58\pm0.04$ for the shell. The total and shell spectral index values fall within the standard range for \acp{SNR} within the \ac{MC}~\citep{Bozzetto2017, 2024MNRAS.529.2443C} and are as expected for an \ac{SNR} predominantly composed of a synchrotron emitting shell. The \ac{PWN} spectral index value is flat, as expected for a \ac{PWN}~\citep{2017Reynolds}. These values differ from previous spectral index values calculated~\citep{klinger2002peculiar}, and this is likely due to the inclusion of more data points than previous estimates, which were obtained from archival radio observations of the \ac{LMC}. Additionally, as we found that the \ac{ATCA} points significantly disagreed with the combined \ac{ATCA} and Parkes data, the inclusion of these outlying points likely impacted the previously calculated values.




\subsection{Surface Brightness}
\label{subsec:surface brightness}

\begin{figure*}[ht!]    
\centering
\includegraphics[width=0.8\textwidth]{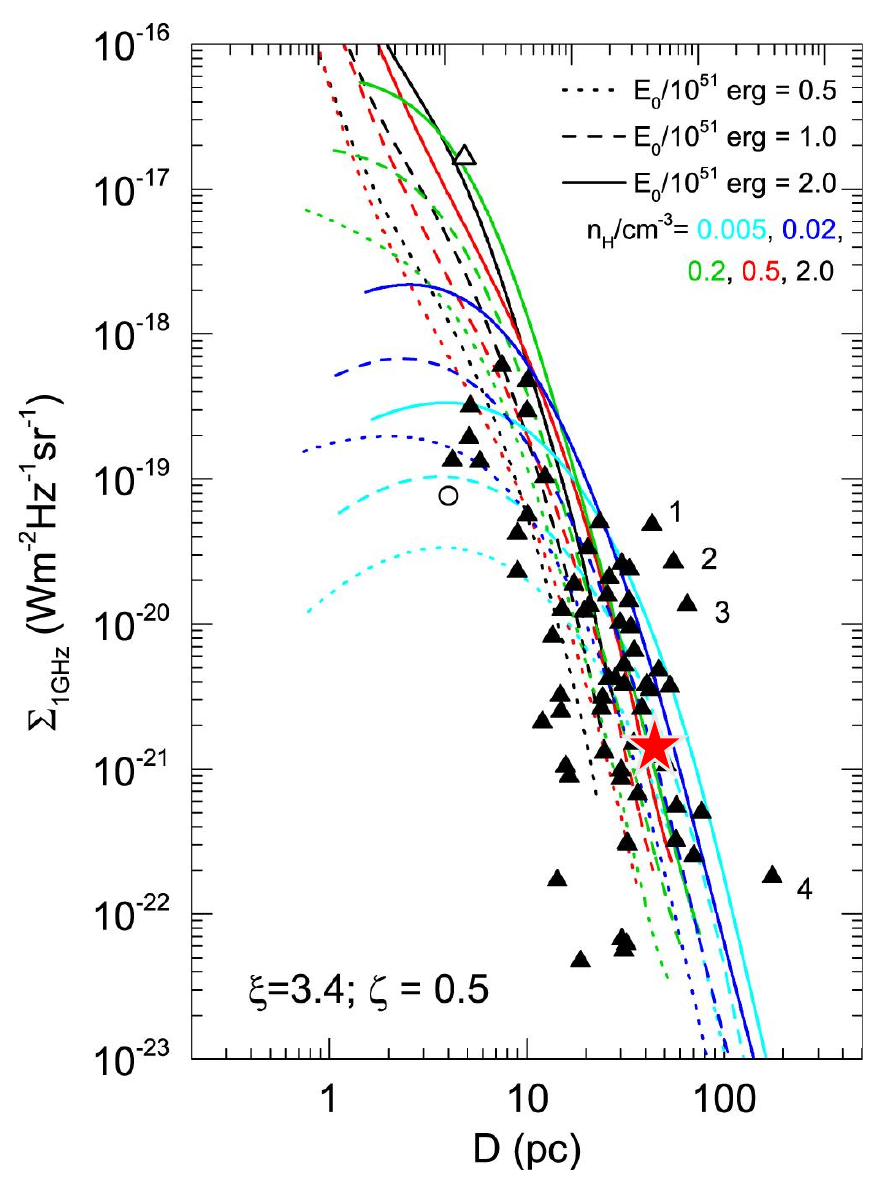}
\caption{Radio surface brightness to diameter diagram for \acp{SNR} at frequency $\nu$\,=\,1\,GHz, adopted from \protect\citet[][their fig.~3]{Pavlovic2018}, shown as black triangles. Different line colours represent different ambient densities, while different line types represent different explosion energies. The open circle is young Galactic \ac{SNR} G1.9+0.3~\citep{Luken2020}, and the open triangle represents Cassiopeia~A. The numbers represent \acp{SNR} (1): CTB~37A, (2): Kes~97, (3): CTB~37B, and (4): G65.1+0.6. The red star represents \snr\ at estimated surface brightness of 1.3$\times$10$^{-21}$\,W\,m$^{-1}$\,Hz$^{-2}$\,sr$^{-1}$ and diameter of 45\,pc (the diameter is taken as the major axis). The image shows evolutionary tracks for representative cases with injection parameter $\xi$\,=\,3.4 and nonlinear magnetic field damping parameter $\zeta$\,=\,0.5.}
\label{fig:sigD}
\end{figure*}

We calculate the radio surface brightness as $\Sigma\,=\,S_{1\,\text{GHz}}/\Omega$ where $\Sigma\,=\,S_{1\,\text{GHz}}$ is the flux density at 1\,GHz (for the shell component), calculated using the measured shell spectral index, and $\Omega$ is the angular area of the source, as defined in Section~\ref{subsec:radio morphology}. We use the shell flux density and spectral index measurements so as to exclude any potential \ac{PWN} contribution and allow a more consistent comparison with other \acp{SNR}. We calculate a surface brightness of 1.3$\times$10$^{-21}$\,W\,m$^{-2}$\,Hz$^{-1}$. We compare these values with the $\Sigma$-D distribution of \citet[their Fig.~3, our Fig.~\ref{fig:sigD}]{Pavlovic2018} and find that \snr\ falls within the typical distribution for \acp{SNR}.

\subsection{Polarisation}
\label{subsec:polarisation}

\begin{figure*}
\centering
\includegraphics[width=0.8\textwidth]{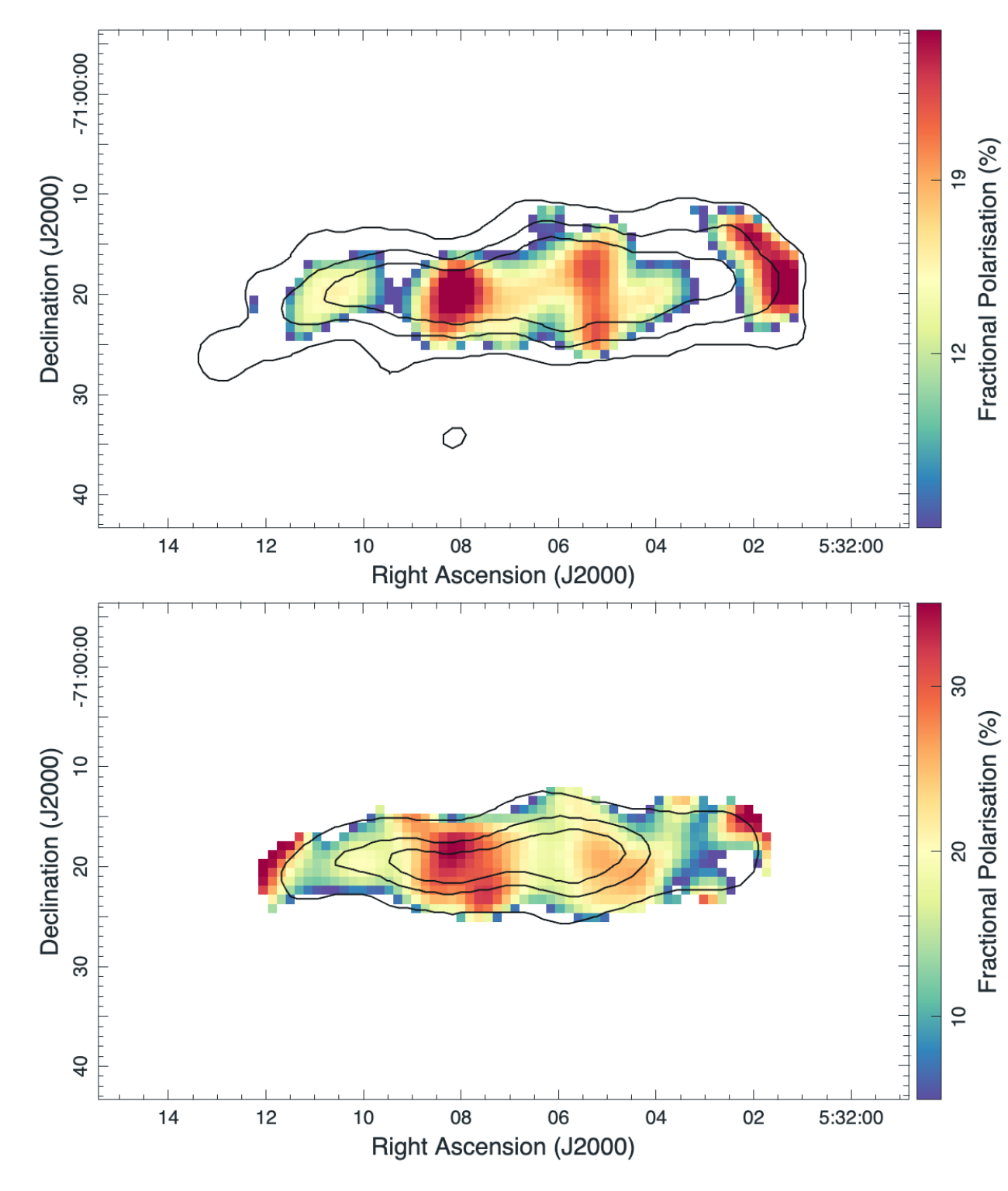}
\caption{Fractional polarisation images of \snr\ \ac{PWN} overlaid with total intensity contours. {\bf Top: } 5500\,MHz fractional polarisation \ac{ATCA} image with 5500\,MHz Stokes~I contours at levels of 0.2, 0.4, and 0.6\,mJy\,beam$^{-1}$. {\bf Bottom: } 9000\,MHz fractional polarisation \ac{ATCA} image with 9000\,MHz Stokes~I contours at levels of 0.2, 0.4, and 0.5\,mJy\,beam$^{-1}$.}
\label{fig:pol}
\end{figure*}

\begin{figure*}[ht!]    

\centering
\includegraphics[ scale=0.57,trim=0 0 0 0,clip]{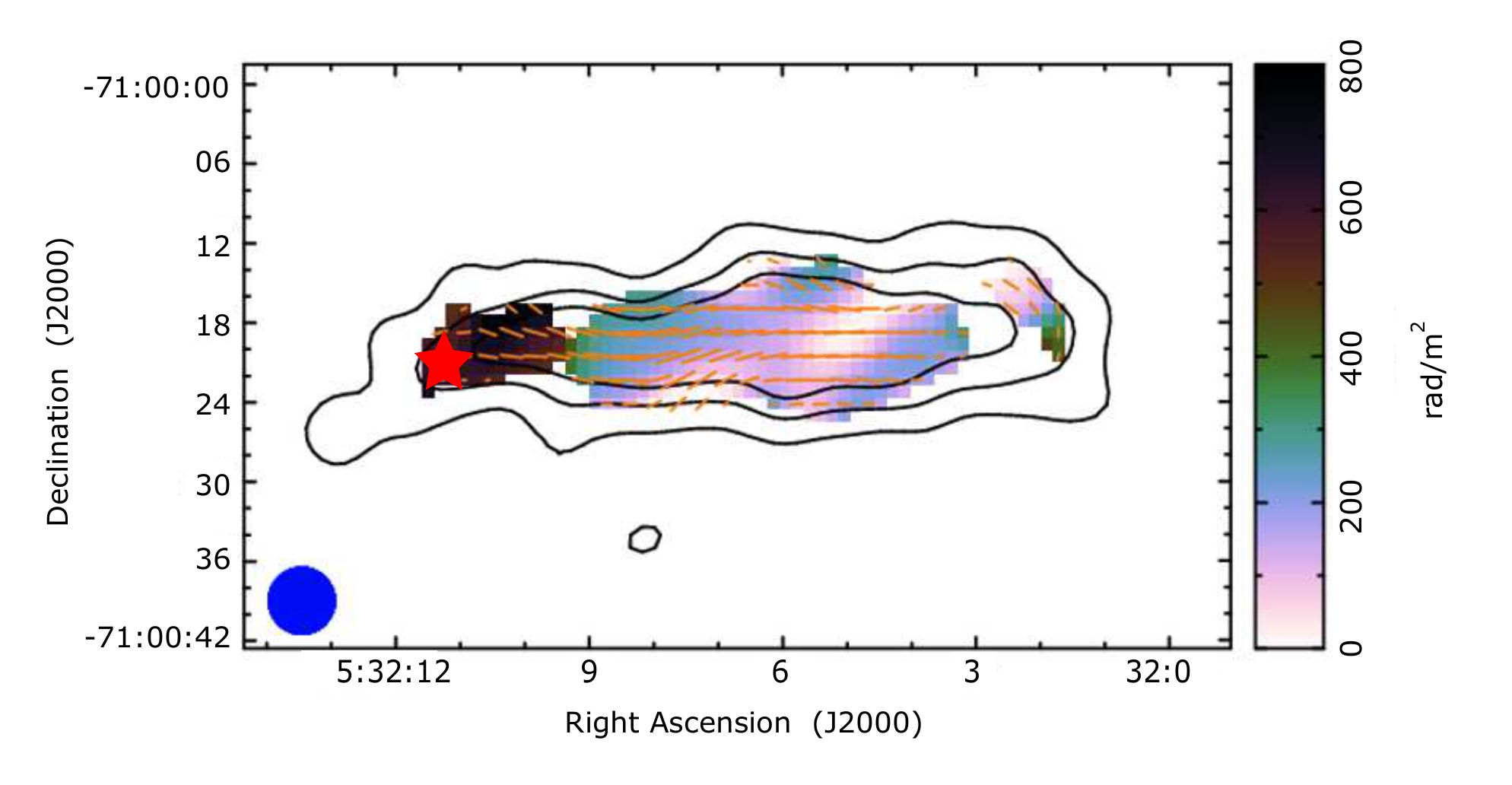}
\caption{$RM$ map calculated between the observations at 5500\,MHz and 9000\,MHz with overlaid vectors in B-field direction, corrected for Faraday rotation. The contour lines are from intensity image at 5500\,MHz at levels of 0.2, 0.4, and 0.6 mJy\,beam$^{-1}$. The red star indicates the location of the X-ray point source.}
\label{fig:MFvec}

\end{figure*}

The \ac{ATCA} observations consisted of Stokes~$Q$ and $U$ observations, allowing us to measure the linear polarisation, rotation measure, and magnetic field of the \ac{PWN}. Due to the lower surface brightness of the \ac{SNR} shell, we were unable to accurately measure the shell's polarisation properties; however, we obtained polarisation measurements for the brighter \ac{PWN} feature. All images were convolved to a common beam size of 5\arcsec$\times$5\arcsec\ to minimise noise, and then the \textsc{miriad} task \textsc{impol} was used to generate the polarised intensity, fractional polarisation, and rotation measure images (see Figs~\ref{fig:pol} and \ref{fig:MFvec}). The measured rotation measure was then used to de-rotate the polarisation vectors to map the electric field vectors, and these were then rotated by 90\D\ to show the intrinsic magnetic field (see Fig.~\ref{fig:MFvec}).

The linear \ac{PWN} feature is visible in both polarisation maps with two areas of higher fractional polarisation in the centre and fading towards the edges. We measure an average fractional polarisation of $3.0\pm1.0$\% with a peak of $9.0\pm1.0$\% at 5.5\,GHz and an average of $3.5\pm2.0$\% with a peak of $26\pm6$\% at 9\,GHz. The rotation measure shows positive values ranging from $\sim$200$-$800 rad m$^{-2}$, with the highest value observed on the eastern tip. This corresponds approximately with the location of the X-ray source~\citep{williams2005supernova}. This may represent the area of a hidden pulsar; however, if this is the case, we would also expect to see a higher fractional polarisation in this area. 

\subsection{Magnetic Field}
\label{subsec:MF}

We observe a highly ordered magnetic field oriented along the linear axis, similar to the results of \citet{klinger2002peculiar}. The ``zig-zag''-like structure observed in the radio-continuum (see Sec.~\ref{subsec:radio morphology}) is also present in the magnetic field structure, with the magnetic field following this pattern. This is expected if this emission represents the tail of a bow-shock \ac{PWN}.

We use an available equipartition model\footnote{\url{http://poincare.matf.bg.ac.rs/~arbo/eqp}}~\citep{2012ApJ...746...79A,2013ApJ...777...31A,2018ApJ...855...59U} to estimate \snr's magnetic field. We use measured values of $\alpha=-0.60$, angular radius $\theta$~=~1\farcm5, $S_{\rm 1\,\text{GHz}}$~=~0.34\,Jy (interpolated from the \ac{EMU} and MeerKAT flux density measurements, see Sec.~\ref{subsec:surface brightness}), and assume a value of $f=0.25$. We find an average magnetic field of $B$\,=\,28.6\,$\mu$G and a minimum energy of $E_\text{min}$\,=\,2.3$\times$10$^{49}$\,ergs when assuming electron equipartition, and values of $B$\,=\,66.3\,$\mu$G and $E_\text{min}$\,=\,1.3$\times$10$^{50}$\,ergs for ion equipartition.

\subsection{''Zig-zag'' pulsar-wind nebula}
\label{subsec:PWN}

The radio-continuum data presented here are consistent with a bow-shock \ac{PWN} scenario. The region exhibits bright radio-continuum emission, accompanied by a flatter radio spectral index, which is consistent with enhanced particle acceleration at this location due to the constant energy input from an energetic pulsar. Similarly, the measured polarisation and ordered magnetic field traces the pulsar's path as it travels supersonically through the \ac{SNR} interior and generates a bow-shock \ac{PWN} behind it. In this scenario, the pulsar would likely be located at the eastern tip of the \ac{PWN} with the tail leading back to the geometric centre of the \ac{SNR}. 

We observe a ''zig-zag''-like structure in the \ac{PWN} which was unseen in previous observations, and has not been observed in other known \ac{PWN} structures. This structure is present in both radio-continuum (see Fig.~\ref{fig:4panels}, top inset) and in the magnetic field vector map (see Fig.~\ref{fig:MFvec}). The fact that it is seen in the radio-continuum images, polarisation images, and the magnetic field indicates that it is an intrinsic physical structure. The zig-zag structure observed corresponds with the brighter areas of polarised emission, which appear almost as polarisation hotspots. 

There are several physical mechanisms which can cause asymmetries or ``knot-like'' structures within highly magnetised plasma structures such as \ac{PWN}. Some of these scenarios could include the formation of kinks, which can appear in some magnetised jets as unstable knots driven by areas of different compression or magnetic instabilities~\citep{2013MNRAS.431L..48P,2019ApJ...884...39B}. Additionally, such a structure may be the result of the \ac{PWN} travelling through the inner \ac{SN} ejecta. Density inhomogeneities within the ejecta may deflect the \ac{PWN} at different points, generating enhanced emission or general asymmetries. Additionally, the pulsar's proper motion may play a part in the structure, for example, if there is significant orbital motion or precession in its motion, then this may generate a ``corkscrew''-like structure. These scenarios are possibilities, and the parameters cannot be constrained until the host pulsar is discovered. 

Another interesting property of the \ac{PWN} is that its direction appears to be aligned with one of the ''ear``-like structures on the eastern side. This is potentially further supported by the \HI\ analysis, where the $p-v$ diagram shows a smaller cavity coincident with the \ac{PWN} structure. We emphasise that this is not a definitive detection however, as the \ac{PWN} size is at the same spatial scale as the \HI\ data resolution, but there is an indication of a possible expanding structure. Therefore, the pulsar may be pushing out of the \ac{SNR} shell, contributing to the observed ``ear'' blowout structure. Additionally, there are some recent theoretical models~\citep{2021MNRAS.502..176C}, which suggest that \ac{SNR} ears may be formed when the \ac{SNR} forward shock interacts with a bipolar circumstellar medium. 

Without further data, all these scenarios are speculative and future observations would be required. What can be seen from the observations is that this zig-zag structure is present, and the most likely cause is some form of asymmetry, interaction, or \ac{PWN} dynamics. Our future high-sensitivity and resolution studies will focus further on this structure to determine its physical origin.

\section{Conclusions}
\label{sec:conclusion}

We have conducted radio-continuum observations and analysis of the \ac{LMC} \ac{SNR} N206, which we have nicknamed ``\snr'' due to the peculiar morphology of its \ac{PWN}. We have focused on the bright \ac{PWN} structure in the east. Specifically, we have calculated a spectral index for the total \ac{SNR} and for the \ac{PWN}, finding a steep ($\alpha\,=\,-0.60$) for the total and flat ($\alpha\,=\,-0.16$) for the \ac{PWN}. These differ from previous spectral index values, likely due to our inclusion of several more data points from archival radio images. This greatly supports the scenario of a typical synchrotron \ac{SNR} shell, with an interior \ac{PWN} with more efficient particle acceleration.

The \ac{PWN} was previously reported as a linear feature; the new observations reveal an unusual "zig-zag" structure, visible in both radio-continuum total intensity images, radio polarisation images, and in the magnetic field structure. The current data do not provide a sufficient explanation for the origin of this structure, but future observations will enable us to better constrain its nature and further our understanding of \ac{PWN} dynamic structures.

\section*{Acknowledgements}

The MeerKAT telescope is operated by the South African Radio Astronomy Observatory, which is a facility of the National Research Foundation, an agency of the Department of Science and Innovation.

This scientific work uses data obtained from Inyarrimanha Ilgari Bundara / the Murchison Radio-astronomy Observatory. We acknowledge the Wajarri Yamaji People as the Traditional Owners and native title holders of the Observatory site. CSIRO’s ASKAP radio telescope is part of the Australia Telescope National Facility\footnote{\label{foot:ATNF}\url{http://www.atnf.csiro.au}}. Operation of ASKAP is funded by the Australian Government with support from the National Collaborative Research Infrastructure Strategy. ASKAP uses the resources of the Pawsey Supercomputing Research Centre. Establishment of ASKAP, Inyarrimanha Ilgari Bundara, the CSIRO Murchison Radio-astronomy Observatory and the Pawsey Supercomputing Research Centre are initiatives of the Australian Government, with support from the Government of Western Australia and the Science and Industry Endowment Fund.

Support for the operation of the MWA is provided by the Australian Government (NCRIS), under a contract to Curtin University administered by Astronomy Australia Limited. 




\newcommand\eprint{in press }

\bibsep=0pt

\bibliographystyle{aa_url_saj}

{\small

\bibliography{N206}

\begin{thebibliography}{110}
\expandafter\ifx\csname natexlab\endcsname\relax\def\natexlab#1{#1}\fi

\bibitem[{{Alsaberi} {et~al.}(2025){Alsaberi}, {Filipovi{\'c}}, {Sano}, {Dai}, {Haberl}, {Kavanagh}, {Leahy}, {Maggi}, {Rowell}, {Sasaki}, {Seitenzahl}, {Uro{\v{s}}evi{\'c}}, {Payne}, {Smeaton}, \& {Lazarevi{\'c}}}]{2025PASA...42...69A}
{Alsaberi}, R. Z.~E., {Filipovi{\'c}}, M.~D., {Sano}, H., {et~al.} 2025, \href{https://ui.adsabs.harvard.edu/abs/2025PASA...42...69A}{\pasa}, \href{https://ui.adsabs.harvard.edu/abs/2025PASA...42...69A}{42, e069}

\bibitem[{{Alsaberi} {et~al.}(2019){Alsaberi}, {Maitra}, {Filipovi{\'c}}, {Bozzetto}, {Haberl}, {Maggi}, {Sasaki}, {Manjolovi{\'c}}, {Velovi{\'c}}, {Kavanagh}, {Maxted}, {Uro{\v{s}}evi{\'c}}, {Rowell}, {Wong}, {For}, {O'Brien}, {Galvin}, {Staveley-Smith}, {Norris}, {Jarrett}, {Kothes}, {Luken}, {Hurley-Walker}, {Sano}, {Oni{\'c}}, {Dai}, {Pannuti}, {Tothill}, {Crawford}, {Yew}, {Boji{\v{c}}i{\'c}}, {D{\'e}nes}, {McClure-Griffiths}, {Gurovich}, \& {Fukui}}]{alsaberi2019discovery}
{Alsaberi}, R. Z.~E., {Maitra}, C., {Filipovi{\'c}}, M.~D., {et~al.} 2019, \href{https://ui.adsabs.harvard.edu/abs/2019MNRAS.486.2507A}{\mnras}, \href{https://ui.adsabs.harvard.edu/abs/2019MNRAS.486.2507A}{486, 2507}

\bibitem[{{Arbutina} {et~al.}(2013){Arbutina}, {Uro{\v s}evi{\'c}}, {Vu{\v c}eti{\'c}}, {Pavlovi{\'c}}, \& {Vukoti{\'c}}}]{2013ApJ...777...31A}
{Arbutina}, B., {Uro{\v s}evi{\'c}}, D., {Vu{\v c}eti{\'c}}, M.~M., {Pavlovi{\'c}}, M.~Z., and {Vukoti{\'c}}, B. 2013, \href{http://adsabs.harvard.edu/abs/2013ApJ...777...31A}{\apj}, \href{http://adsabs.harvard.edu/abs/2013ApJ...777...31A}{777, 31}

\bibitem[{{Arbutina} {et~al.}(2012){Arbutina}, {Uro{\v{s}}evi{\'c}}, {Andjeli{\'c}}, {Pavlovi{\'c}}, \& {Vukoti{\'c}}}]{2012ApJ...746...79A}
{Arbutina}, B., {Uro{\v{s}}evi{\'c}}, D., {Andjeli{\'c}}, M.~M., {Pavlovi{\'c}}, M.~Z., and {Vukoti{\'c}}, B. 2012, \href{https://ui.adsabs.harvard.edu/abs/2012ApJ...746...79A}{\apj}, \href{https://ui.adsabs.harvard.edu/abs/2012ApJ...746...79A}{746, 79}

\bibitem[{{Asher} {et~al.}(2025){Asher}, {Smeaton}, {Filipovi{\'c}}, {Hopkins}, {van Loon}, {Galvin}, \& {Barnes}}]{2025arXiv250606768A}
{Asher}, A.~D., {Smeaton}, Z.~J., {Filipovi{\'c}}, M.~D., {et~al.} 2025, {arXiv e-prints, \href{https://ui.adsabs.harvard.edu/abs/2025arXiv250606768A}{}\eprint  \href{https://arxiv.org/abs/2506.06768}{(DOI:  )}, arXiv:2506.06768}

\bibitem[{{Bordiu} {et~al.}(2024){Bordiu}, {Filipovi{\'c}}, {Umana}, {Cotton}, {Buemi}, {Bufano}, {Camilo}, {Cavallaro}, {Cerrigone}, {Dai}, {Hopkins}, {Ingallinera}, {Jarrett}, {Koribalski}, {Lazarevi{\'c}}, {Leto}, {Loru}, {Lundqvist}, {Mackey}, {Norris}, {Payne}, {Rowell}, {Riggi}, {Rizzo}, {Ruggeri}, {Shabala}, {Smeaton}, {Trigilio}, \& {Velovi{\'c}}}]{2024A&A...690A..53B}
{Bordiu}, C., {Filipovi{\'c}}, M.~D., {Umana}, G., {et~al.} 2024, \href{https://ui.adsabs.harvard.edu/abs/2024A&A...690A..53B}{\aap}, \href{https://ui.adsabs.harvard.edu/abs/2024A&A...690A..53B}{690, A53}

\bibitem[{{Bozzetto} and {Filipovi{\'c}}(2014){Bozzetto} \& {Filipovi{\'c}}}]{2014Ap&SS.351..207B}
{Bozzetto}, L.~M. and {Filipovi{\'c}}, M.~D. 2014, \href{https://ui.adsabs.harvard.edu/abs/2014Ap&SS.351..207B}{\apss}, \href{https://ui.adsabs.harvard.edu/abs/2014Ap&SS.351..207B}{351, 207}

\bibitem[{{Bozzetto} {et~al.}(2012){Bozzetto}, {Filipovic}, {Crawford}, {De Horta}, \& {Stupar}}]{2012SerAJ.184...69B}
{Bozzetto}, L.~M., {Filipovic}, M.~D., {Crawford}, E.~J., {De Horta}, A.~Y., and {Stupar}, M. 2012, \href{https://ui.adsabs.harvard.edu/abs/2012SerAJ.184...69B}{Serbian Astronomical Journal}, \href{https://ui.adsabs.harvard.edu/abs/2012SerAJ.184...69B}{184, 69}

\bibitem[{{Bozzetto} {et~al.}(2015){Bozzetto}, {Filipovic}, {Haberl}, {Sasaki}, {Kavanagh}, {Maggi}, {Urosevic}, \& {Sturm}}]{2015PKAS...30..149B}
{Bozzetto}, L.~M., {Filipovic}, M.~D., {Haberl}, F., {et~al.} 2015, \href{https://ui.adsabs.harvard.edu/abs/2015PKAS...30..149B}{Publication of Korean Astronomical Society}, \href{https://ui.adsabs.harvard.edu/abs/2015PKAS...30..149B}{30, 149}

\bibitem[{{Bozzetto} {et~al.}(2023){Bozzetto}, {Filipovi{\'c}}, {Sano}, {Alsaberi}, {Barnes}, {Boji{\v{c}}i{\'c}}, {Brose}, {Chomiuk}, {Crawford}, {Dai}, {Ghavam}, {Haberl}, {Hill}, {Hopkins}, {Ingallinera}, {Jarrett}, {Kavanagh}, {Koribalski}, {Kothes}, {Leahy}, {Lenc}, {Leonidaki}, {Maggi}, {Maitra}, {Matthew}, {Payne}, {Pennock}, {Points}, {Reid}, {Riggi}, {Rowell}, {Sasaki}, {Safi-Harb}, {van Loon}, {Tothill}, {Uro{\v{s}}evi{\'c}}, \& {Zangrandi}}]{Bozzetto2023}
{Bozzetto}, L.~M., {Filipovi{\'c}}, M.~D., {Sano}, H., {et~al.} 2023, \href{https://ui.adsabs.harvard.edu/abs/2023MNRAS.518.2574B}{\mnras}, \href{https://ui.adsabs.harvard.edu/abs/2023MNRAS.518.2574B}{518, 2574}

\bibitem[{{Bozzetto} {et~al.}(2017){Bozzetto}, {Filipovi{\'c}}, {Vukoti{\'c}}, {Pavlovi{\'c}}, {Uro{\v{s}}evi{\'c}}, {Kavanagh}, {Arbutina}, {Maggi}, {Sasaki}, {Haberl}, {Crawford}, {Roper}, {Grieve}, \& {Points}}]{Bozzetto2017}
{Bozzetto}, L.~M., {Filipovi{\'c}}, M.~D., {Vukoti{\'c}}, B., {et~al.} 2017, \href{https://ui.adsabs.harvard.edu/abs/2017ApJS..230....2B}{\apjs}, \href{https://ui.adsabs.harvard.edu/abs/2017ApJS..230....2B}{230, 2}

\bibitem[{{Bradley} {et~al.}(2025){Bradley}, {Smeaton}, {Tothill}, {Filipovi{\'c}}, {Becker}, {Hopkins}, {Koribalski}, {Lazarevi{\'c}}, {Leahy}, {Rowell}, {Velovi{\'c}}, \& {Uro{\v{s}}evi{\'c}}}]{Bradley2025}
{Bradley}, A., {Smeaton}, Z., {Tothill}, N., {et~al.} 2025, \href{https://ui.adsabs.harvard.edu/abs/2025PASA...42...32B}{\pasa}, \href{https://ui.adsabs.harvard.edu/abs/2025PASA...42...32B}{42, e032}

\bibitem[{{Brantseg} {et~al.}(2014){Brantseg}, {McEntaffer}, {Bozzetto}, {Filipovic}, \& {Grieves}}]{2014ApJ...780...50B}
{Brantseg}, T., {McEntaffer}, R.~L., {Bozzetto}, L.~M., {Filipovic}, M., and {Grieves}, N. 2014, \href{https://ui.adsabs.harvard.edu/abs/2014ApJ...780...50B}{\apj}, \href{https://ui.adsabs.harvard.edu/abs/2014ApJ...780...50B}{780, 50}

\bibitem[{{Bromberg} {et~al.}(2019){Bromberg}, {Singh}, {Davelaar}, \& {Philippov}}]{2019ApJ...884...39B}
{Bromberg}, O., {Singh}, C.~B., {Davelaar}, J., and {Philippov}, A.~A. 2019, \href{https://ui.adsabs.harvard.edu/abs/2019ApJ...884...39B}{\apj}, \href{https://ui.adsabs.harvard.edu/abs/2019ApJ...884...39B}{884, 39}

\bibitem[{{Burger-Scheidlin} {et~al.}(2024){Burger-Scheidlin}, {Brose}, {Mackey}, {Filipovi{\'c}}, {Goswami}, {Guillen}, {de O{\~n}a Wilhelmi}, \& {Sushch}}]{BurgerSchiedlin2024}
{Burger-Scheidlin}, C., {Brose}, R., {Mackey}, J., {et~al.} 2024, \href{https://ui.adsabs.harvard.edu/abs/2024A&A...684A.150B}{A\&A}, \href{https://ui.adsabs.harvard.edu/abs/2024A&A...684A.150B}{684, A150}

\bibitem[{{Camilo} {et~al.}(2002){Camilo}, {Manchester}, {Gaensler}, \& {Lorimer}}]{2002ApJ...579L..25C}
{Camilo}, F., {Manchester}, R.~N., {Gaensler}, B.~M., and {Lorimer}, D.~R. 2002, \href{https://ui.adsabs.harvard.edu/abs/2002ApJ...579L..25C}{\apjl}, \href{https://ui.adsabs.harvard.edu/abs/2002ApJ...579L..25C}{579, L25}

\bibitem[{{Chiotellis} {et~al.}(2021){Chiotellis}, {Boumis}, \& {Spetsieri}}]{2021MNRAS.502..176C}
{Chiotellis}, A., {Boumis}, P., and {Spetsieri}, Z.~T. 2021, \href{https://ui.adsabs.harvard.edu/abs/2021MNRAS.502..176C}{\mnras}, \href{https://ui.adsabs.harvard.edu/abs/2021MNRAS.502..176C}{502, 176}

\bibitem[{Comrie {et~al.}(2018)Comrie, Wang, Hsu, Moraghan, Harris, Pang, Pińska, Chiang, Simmonds, Chang, Hwang, Jan, \& Lin}]{CARTA_2018}
Comrie, A., Wang, K.-S., Hsu, S.-C., {et~al.} 2018, \href{}{CARTA: The Cube Analysis and Rendering Tool for Astronomy}

\bibitem[{{Cotton} {et~al.}(2024){Cotton}, {Filipovi{\'c}}, {Camilo}, {Indebetouw}, {Alsaberi}, {Anih}, {Baker}, {Bastian}, {Boji{\v{c}}i{\'c}}, {Carli}, {Cavallaro}, {Crawford}, {Dai}, {Haberl}, {Levin}, {Luken}, {Pennock}, {Rajabpour}, {Stappers}, {van Loon}, {Zijlstra}, {Buchner}, {Geyer}, {Goedhart}, \& {Serylak}}]{2024MNRAS.529.2443C}
{Cotton}, W.~D., {Filipovi{\'c}}, M.~D., {Camilo}, F., {et~al.} 2024, \href{https://ui.adsabs.harvard.edu/abs/2024MNRAS.529.2443C}{\mnras}, \href{https://ui.adsabs.harvard.edu/abs/2024MNRAS.529.2443C}{529, 2443}

\bibitem[{{Crawford} {et~al.}(2008){Crawford}, {Filipovi\'c}, {de Horta}, {Stootman}, \& {Payne}}]{Crawford2008}
{Crawford}, E.~J., {Filipovi\'c}, M.~D., {de Horta}, A.~Y., {Stootman}, F.~H., and {Payne}, J.~L. 2008, \href{https://ui.adsabs.harvard.edu/abs/2008SerAJ.177...61C}{Serbian Astronomical Journal}, \href{https://ui.adsabs.harvard.edu/abs/2008SerAJ.177...61C}{177, 61}

\bibitem[{{Dickel} {et~al.}(2005){Dickel}, {McIntyre}, {Gruendl}, \& {Milne}}]{2005AJ....129..790D}
{Dickel}, J.~R., {McIntyre}, V.~J., {Gruendl}, R.~A., and {Milne}, D.~K. 2005, \href{https://ui.adsabs.harvard.edu/abs/2005AJ....129..790D}{\aj}, \href{https://ui.adsabs.harvard.edu/abs/2005AJ....129..790D}{129, 790}

\bibitem[{{Dimaratos} {et~al.}(2015){Dimaratos}, {Cormier}, {Bigiel}, \& {Madden}}]{Dimaratos2015}
{Dimaratos}, A., {Cormier}, D., {Bigiel}, F., and {Madden}, S.~C. 2015, \href{https://ui.adsabs.harvard.edu/abs/2015A&A...580A.135D}{\aap}, \href{https://ui.adsabs.harvard.edu/abs/2015A&A...580A.135D}{580, A135}

\bibitem[{{Filipovi{\'c}} {et~al.}(2021){Filipovi{\'c}}, {Boji{\v{c}}i{\'c}}, {Grieve}, {Norris}, {Tothill}, {Shobhana}, {Rudnick}, {Prandoni}, {Andernach}, {Hurley-Walker}, {Alsaberi}, {Anderson}, {Collier}, {Crawford}, {For}, {Galvin}, {Haberl}, {Hopkins}, {Ingallinera}, {Kavanagh}, {Koribalski}, {Kothes}, {Leahy}, {Leverenz}, {Maggi}, {Maitra}, {Marvil}, {Pannuti}, {Park}, {Payne}, {Pennock}, {Riggi}, {Rowell}, {Sano}, {Sasaki}, {Staveley-Smith}, {Trigilio}, {Umana}, {Uro{\v{s}}evi{\'c}}, {van Loon}, \& {Vardoulaki}}]{2021MNRAS.507.2885F}
{Filipovi{\'c}}, M.~D., {Boji{\v{c}}i{\'c}}, I.~S., {Grieve}, K.~R., {et~al.} 2021, \href{https://ui.adsabs.harvard.edu/abs/2021MNRAS.507.2885F}{\mnras}, \href{https://ui.adsabs.harvard.edu/abs/2021MNRAS.507.2885F}{507, 2885}

\bibitem[{{Filipovi{\'c}} {et~al.}(2023){Filipovi{\'c}}, {Dai}, {Arbutina}, {Hurley-Walker}, {Brose}, {Becker}, {Sano}, {Uro{\v{s}}evi{\'c}}, {Jarrett}, {Hopkins}, {Alsaberi}, {Alsulami}, {Bordiu}, {Ball}, {Bufano}, {Burger-Scheidlin}, {Crawford}, {English}, {Haberl}, {Ingallinera}, {Kapinska}, {Kavanagh}, {Koribalski}, {Kothes}, {Lazarevi{\'c}}, {Mackey}, {Rowell}, {Leahy}, {Loru}, {Macgregor}, {Nicastro}, {Norris}, {Riggi}, {Sasaki}, {Stupar}, {Trigilio}, {Umana}, {Vernstrom}, \& {Vukoti{\'c}}}]{Filipovic2023}
{Filipovi{\'c}}, M.~D., {Dai}, S., {Arbutina}, B., {et~al.} 2023, \href{https://ui.adsabs.harvard.edu/abs/2023AJ....166..149F}{\aj}, \href{https://ui.adsabs.harvard.edu/abs/2023AJ....166..149F}{166, 149}

\bibitem[{{Filipovi\'c} {et~al.}(1998{\natexlab{a}}){Filipovi\'c}, {Haynes}, {White}, \& {Jones}}]{Filipovic1998a}
{Filipovi\'c}, M.~D., {Haynes}, R.~F., {White}, G.~L., and {Jones}, P.~A. 1998{\natexlab{a}}, \href{https://ui.adsabs.harvard.edu/abs/1998A&AS..130..421F}{\aaps}, \href{https://ui.adsabs.harvard.edu/abs/1998A&AS..130..421F}{130, 421}

\bibitem[{{Filipovi\'c} {et~al.}(1998{\natexlab{b}}){Filipovi\'c}, {Haynes}, {White}, \& {Jones}}]{Filipovic1998PaperVII}
{Filipovi\'c}, M.~D., {Haynes}, R.~F., {White}, G.~L., and {Jones}, P.~A. 1998{\natexlab{b}}, \href{https://ui.adsabs.harvard.edu/abs/1998A&AS..130..421F}{\aaps}, \href{https://ui.adsabs.harvard.edu/abs/1998A&AS..130..421F}{130, 421}

\bibitem[{{Filipovi\'c} {et~al.}(1995){Filipovi\'c}, {Haynes}, {White}, {Jones}, {Klein}, \& {Wielebinski}}]{Filipovic1995PaperIV}
{Filipovi\'c}, M.~D., {Haynes}, R.~F., {White}, G.~L., {et~al.} 1995, \href{https://ui.adsabs.harvard.edu/abs/1995A&AS..111..311F}{\aaps}, \href{https://ui.adsabs.harvard.edu/abs/1995A&AS..111..311F}{111, 311}

\bibitem[{{Filipovi{\'c}} {et~al.}(2024){Filipovi{\'c}}, {Lazarevi{\'c}}, {Araya}, {Hurley-Walker}, {Kothes}, {Sano}, {Rowell}, {Martin}, {Fukui}, {Alsaberi}, {Arbutina}, {Ball}, {Bordiu}, {Brose}, {Bufano}, {Burger-Scheidlin}, {Anne Collins}, {Crawford}, {Dai}, {William Duchesne}, {Fuller}, {Hopkins}, {Ingallinera}, {Inoue}, {Jarrett}, {Koribalski}, {Leahy}, {Luken}, {Mackey}, {Macgregor}, {Norris}, {Payne}, {Riggi}, {Riseley}, {Sasaki}, {Smeaton}, {Sushch}, {Stupar}, {Umana}, {Uro{\v{s}}evi{\'c}}, {Velovi{\'c}}, {Vernstrom}, {Vukoti{\'c}}, \& {West}}]{Diprotodon}
{Filipovi{\'c}}, M.~D., {Lazarevi{\'c}}, S., {Araya}, M., {et~al.} 2024, \href{https://ui.adsabs.harvard.edu/abs/2024PASA...41..112F}{\pasa}, \href{https://ui.adsabs.harvard.edu/abs/2024PASA...41..112F}{41, e112}

\bibitem[{{Filipovi{\'c}} {et~al.}(2022){Filipovi{\'c}}, {Payne}, {Alsaberi}, {Norris}, {Macgregor}, {Rudnick}, {Koribalski}, {Leahy}, {Ducci}, {Kothes}, {Andernach}, {Barnes}, {Boji{\v{c}}i{\'c}}, {Bozzetto}, {Brose}, {Collier}, {Crawford}, {Crocker}, {Dai}, {Galvin}, {Haberl}, {Heber}, {Hill}, {Hopkins}, {Hurley-Walker}, {Ingallinera}, {Jarrett}, {Kavanagh}, {Lenc}, {Luken}, {Mackey}, {Manojlovi{\'c}}, {Maggi}, {Maitra}, {Pennock}, {Points}, {Riggi}, {Rowell}, {Safi-Harb}, {Sano}, {Sasaki}, {Shabala}, {Stevens}, {van Loon}, {Tothill}, {Umana}, {Uro{\v{s}}evi{\'c}}, {Velovi{\'c}}, {Vernstrom}, {West}, \& {Wan}}]{2022MNRAS.512..265F}
{Filipovi{\'c}}, M.~D., {Payne}, J.~L., {Alsaberi}, R.~Z.~E., {et~al.} 2022, \href{https://ui.adsabs.harvard.edu/abs/2022MNRAS.512..265F}{\mnras}, \href{https://ui.adsabs.harvard.edu/abs/2022MNRAS.512..265F}{512, 265}

\bibitem[{{Filipovi{\'c}} {et~al.}(2025){Filipovi{\'c}}, {Smeaton}, {Bradley}, {Dobie}, {Koribalski}, {Kothes}, {Rudnick}, {Ahmad}, {Alsaberi}, {Anderson}, {Barnes}, {Breuhaus}, {Crawford}, {Dai}, {Gordon}, {Gupta}, {Hopkins}, {Leahy}, {Luken}, {McClure-Griffiths}, {Micha{\l}owski}, {Sasaki}, {Tothill}, {Umana}, {Vernstrom}, \& {West}}]{2025ApJ...984L..52F}
{Filipovi{\'c}}, M.~D., {Smeaton}, Z.~J., {Bradley}, A.~C., {et~al.} 2025, \href{https://ui.adsabs.harvard.edu/abs/2025ApJ...984L..52F}{\apjl}, \href{https://ui.adsabs.harvard.edu/abs/2025ApJ...984L..52F}{984, L52}

\bibitem[{{Filipovic} {et~al.}(2025){Filipovic}, {Smeaton}, {Kothes}, {Mantovanini}, {Kostic}, {Leahy}, {Ahmad}, {Anderson}, {Araya}, {Ball}, {Becker}, {Bordiu}, {Bradley}, {Brose}, {Burger-Scheidlin}, {Dai}, {Duchesne}, {Galvin}, {Hopkins}, {Hurley-Walker}, {Koribalski}, {Lazarevic}, {Lundqvist}, {Mackey}, {Martin}, {McGee}, {Mitrasinovic}, {Payne}, {Riggi}, {Ross}, {Rowell}, {Rudnick}, {Sano}, {Sasaki}, {Soria}, {Urosevic}, {Vukotic}, \& {West}}]{2025arXiv250504041F}
{Filipovic}, M.~D., {Smeaton}, Z.~J., {Kothes}, R., {et~al.} 2025, {arXiv e-prints, \href{https://ui.adsabs.harvard.edu/abs/2025arXiv250504041F}{}\eprint  \href{https://arxiv.org/abs/2505.04041}{(DOI:  10.48550/arXiv.2505.04041)}, arXiv:2505.04041}

\bibitem[{{Filipovi{\'c}} and {Tothill}(2021){Filipovi{\'c}} \& {Tothill}}]{2021pma..book.....F}
{Filipovi{\'c}}, M.~D. and {Tothill}, N. F.~H. 2021, {Principles of Multimessenger Astronomy}, AAS-IOP astronomy (Institute of Physics Publishing)

\bibitem[{{Filipovic} {et~al.}(1996){Filipovic}, {White}, {Jones}, {Haynes}, {Pietsch}, {Wielebinski}, \& {Klein}}]{1996ASPC..112...91F}
{Filipovic}, M.~D., {White}, G.~L., {Jones}, P.~A., {et~al.} 1996, in Astronomical Society of the Pacific Conference Series, Vol. 112, The History of the Milky Way and Its Satellite System, ed. A.~{Burkert}, D.~H. {Hartmann}, and S.~A. {Majewski}, 91

\bibitem[{{For} {et~al.}(2018){For}, {Staveley-Smith}, {Hurley-Walker}, {Franzen}, {Kapi{\'n}ska}, {Filipovi{\'c}}, {Collier}, {Wu}, {Grieve}, {Callingham}, {Bell}, {Bernardi}, {Bowman}, {Briggs}, {Cappallo}, {Deshpande}, {Dwarakanath}, {Gaensler}, {Greenhill}, {Hancock}, {Hazelton}, {Hindson}, {Johnston-Hollitt}, {Kaplan}, {Lenc}, {Lonsdale}, {McKinley}, {McWhirter}, {Mitchell}, {Morales}, {Morgan}, {Morgan}, {Oberoi}, {Offringa}, {Ord}, {Prabu}, {Procopio}, {Shankar}, {Srivani}, {Subrahmanyan}, {Tingay}, {Wayth}, {Webster}, {Williams}, {Williams}, \& {Zheng}}]{2018MNRAS.480.2743F}
{For}, B.-Q., {Staveley-Smith}, L., {Hurley-Walker}, N., {et~al.} 2018, \href{http://adsabs.harvard.edu/abs/2018MNRAS.480.2743F}{\mnras}, \href{http://adsabs.harvard.edu/abs/2018MNRAS.480.2743F}{480, 2743}

\bibitem[{{Fukui} {et~al.}(2024){Fukui}, {Aruga}, {Sano}, {Hayakawa}, {Inoue}, {Rowell}, {Einecke}, \& {Tachihara}}]{2024ApJ...961..162F}
{Fukui}, Y., {Aruga}, M., {Sano}, H., {et~al.} 2024, \href{https://ui.adsabs.harvard.edu/abs/2024ApJ...961..162F}{\apj}, \href{https://ui.adsabs.harvard.edu/abs/2024ApJ...961..162F}{961, 162}

\bibitem[{{Ghavam} {et~al.}(2024){Ghavam}, {Filipovi{\'c}}, {Alsaberi}, {Barnes}, {Crawford}, {Haberl}, {Kavanagh}, {Maggi}, {Payne}, {Rowell}, {Hidetoshi}, {Sasaki}, {Rajabpour}, {Tothill}, \& {Uro{\v{s}}evi{\'c}}}]{2024PASA...41...89G}
{Ghavam}, M., {Filipovi{\'c}}, M.~D., {Alsaberi}, R., {et~al.} 2024, \href{https://ui.adsabs.harvard.edu/abs/2024PASA...41...89G}{\pasa}, \href{https://ui.adsabs.harvard.edu/abs/2024PASA...41...89G}{41, e089}

\bibitem[{{Gooch}(1995)}]{1995Gooch}
{Gooch}, R. 1995, in Astronomical Society of the Pacific Conference Series, Vol.~77, Astronomical Data Analysis Software and Systems IV, ed. R.~A. {Shaw}, H.~E. {Payne}, and J.~J.~E. {Hayes}, 144

\bibitem[{{Gorjian} {et~al.}(2004){Gorjian}, {Werner}, {Mould}, {Gordon}, {Muzzerole}, {Morrison}, {Surace}, {Rebull}, {Hurt}, {Smith}, {Points}, {Aguilera}, {De Buizer}, \& {Packham}}]{2004ApJS..154..275G}
{Gorjian}, V., {Werner}, M.~W., {Mould}, J.~R., {et~al.} 2004, \href{https://ui.adsabs.harvard.edu/abs/2004ApJS..154..275G}{\apjs}, \href{https://ui.adsabs.harvard.edu/abs/2004ApJS..154..275G}{154, 275}

\bibitem[{{Guzman} {et~al.}(2019){Guzman}, {Whiting}, {Voronkov}, {Mitchell}, {Ord}, {Collins}, {Marquarding}, {Lahur}, {Maher}, {Van Diepen}, {Bannister}, {Wu}, {Lenc}, {Khoo}, \& {Bastholm}}]{Guzman2019}
{Guzman}, J., {Whiting}, M., {Voronkov}, M., {et~al.} 2019, \href{https://ui.adsabs.harvard.edu/abs/2019ascl.soft12003G}{{ASKAPsoft: ASKAP science data processor software}}

\bibitem[{{Haberl}(2014)}]{Haberl2014}
{Haberl}, F. 2014, in The X-ray Universe 2014, ed. J.-U. {Ness}, 4

\bibitem[{{Haberl} {et~al.}(2012){Haberl}, {Filipovi{\'c}}, {Bozzetto}, {Crawford}, {Points}, {Pietsch}, {De Horta}, {Tothill}, {Payne}, \& {Sasaki}}]{2012Harbel}
{Haberl}, F., {Filipovi{\'c}}, M.~D., {Bozzetto}, L.~M., {et~al.} 2012, \href{https://ui.adsabs.harvard.edu/\#abs/2012A&A...543A.154H}{\aap}, \href{https://ui.adsabs.harvard.edu/\#abs/2012A&A...543A.154H}{543, A154}

\bibitem[{{Hales} {et~al.}(2009){Hales}, {Gaensler}, {Chatterjee}, {van der Swaluw}, \& {Camilo}}]{2009ApJ...706.1316H}
{Hales}, C.~A., {Gaensler}, B.~M., {Chatterjee}, S., {van der Swaluw}, E., and {Camilo}, F. 2009, \href{https://ui.adsabs.harvard.edu/abs/2009ApJ...706.1316H}{\apj}, \href{https://ui.adsabs.harvard.edu/abs/2009ApJ...706.1316H}{706, 1316}

\bibitem[{{Henize}(1956)}]{Henize1956}
{Henize}, K.~G. 1956, \href{https://ui.adsabs.harvard.edu/abs/1956ApJS....2..315H}{\apjs}, \href{https://ui.adsabs.harvard.edu/abs/1956ApJS....2..315H}{2, 315}

\bibitem[{{Hopkins} {et~al.}(2025){Hopkins}, {Kapinska}, {Marvil}, {Vernstrom}, {Collier}, {Norris}, {Gordon}, {Duchesne}, {Rudnick}, {Gupta}, {Carretti}, {Anderson}, {Dai}, {G{\"u}rkan}, {Parkinson}, {Prandoni}, {Riggi}, {Shekhar Saraf}, {Ma}, {Filipovi{\'c}}, {Umana}, {Bahr-Kalus}, {Koribalski}, {Lenc}, {Ingallinera}, {Afonso}, {Ahmad}, {Ahmed}, {Alexander}, {Andernach}, {Asorey}, {Battisti}, {Bilicki}, {Botteon}, {Brown}, {Br{\"u}ggen}, {Cowley}, {Dage}, {Hale}, {Hardcastle}, {Kothes}, {Lazarevi{\'c}}, {Lin}, {Luken}, {Moss}, {Prathap}, {ur Rahman}, {Reiprich}, {Riseley}, {Salvato}, {Seymour}, {Shabala}, {Smith}, {Vaccari}, {van Loon}, {Wong}, {Zainal Alsaberi}, {Asher}, {Ball}, {Barbosa}, {Biava}, {Bradley}, {Carvajal}, {Crawford}, {Galvin}, {Huynh}, {Leahy}, {Matute}, {Moss}, {Pappalardo}, {Smeaton}, {Velovi{\'c}}, \& {Zafar}}]{2025PASA...42...71H}
{Hopkins}, A., {Kapinska}, A., {Marvil}, J., {et~al.} 2025, \href{https://ui.adsabs.harvard.edu/abs/2025PASA...42...71H}{\pasa}, \href{https://ui.adsabs.harvard.edu/abs/2025PASA...42...71H}{42, e071}

\bibitem[{{Hughes} {et~al.}(2007){Hughes}, {Staveley-Smith}, {Kim}, {Wolleben}, \& {Filipovi{\'c}}}]{2007MNRAS.382..543H}
{Hughes}, A., {Staveley-Smith}, L., {Kim}, S., {Wolleben}, M., and {Filipovi{\'c}}, M. 2007, \href{https://ui.adsabs.harvard.edu/abs/2007MNRAS.382..543H}{\mnras}, \href{https://ui.adsabs.harvard.edu/abs/2007MNRAS.382..543H}{382, 543}

\bibitem[{{Hurley-Walker} {et~al.}(2017){Hurley-Walker}, {Callingham}, {Hancock}, {Franzen}, {Hindson}, {Kapi{\'n}ska}, {Morgan}, {Offringa}, {Wayth}, {Wu}, {Zheng}, {Murphy}, {Bell}, {Dwarakanath}, {For}, {Gaensler}, {Johnston-Hollitt}, {Lenc}, {Procopio}, {Staveley-Smith}, {Ekers}, {Bowman}, {Briggs}, {Cappallo}, {Deshpande}, {Greenhill}, {Hazelton}, {Kaplan}, {Lonsdale}, {McWhirter}, {Mitchell}, {Morales}, {Morgan}, {Oberoi}, {Ord}, {Prabu}, {Shankar}, {Srivani}, {Subrahmanyan}, {Tingay}, {Webster}, {Williams}, \& {Williams}}]{HurleyWalker2017}
{Hurley-Walker}, N., {Callingham}, J.~R., {Hancock}, P.~J., {et~al.} 2017, \href{https://ui.adsabs.harvard.edu/abs/2017MNRAS.464.1146H}{\mnras}, \href{https://ui.adsabs.harvard.edu/abs/2017MNRAS.464.1146H}{464, 1146}

\bibitem[{{Hurley-Walker} {et~al.}(2019){Hurley-Walker}, {Gaensler}, {Leahy}, {Filipovi{\'c}}, {Hancock}, {Franzen}, {Offringa}, {Callingham}, {Hindson}, {Wu}, {Bell}, {For}, {Johnston-Hollitt}, {Kapi{\'n}ska}, {Morgan}, {Murphy}, {McKinley}, {Procopio}, {Staveley-Smith}, {Wayth}, \& {Zheng}}]{2019PASA...36...48H}
{Hurley-Walker}, N., {Gaensler}, B.~M., {Leahy}, D.~A., {et~al.} 2019, \href{https://ui.adsabs.harvard.edu/abs/2019PASA...36...48H}{\pasa}, \href{https://ui.adsabs.harvard.edu/abs/2019PASA...36...48H}{36, e048}

\bibitem[{{Kavanagh} {et~al.}(2012){Kavanagh}, {Sasaki}, \& {Points}}]{2012A&A...547A..19K}
{Kavanagh}, P.~J., {Sasaki}, M., and {Points}, S.~D. 2012, \href{https://ui.adsabs.harvard.edu/abs/2012A&A...547A..19K}{\aap}, \href{https://ui.adsabs.harvard.edu/abs/2012A&A...547A..19K}{547, A19}

\bibitem[{{Kim} {et~al.}(2003){Kim}, {Staveley-Smith}, {Dopita}, {Sault}, {Freeman}, {Lee}, \& {Chu}}]{2003ApJS..148..473K}
{Kim}, S., {Staveley-Smith}, L., {Dopita}, M.~A., {et~al.} 2003, \href{https://ui.adsabs.harvard.edu/abs/2003ApJS..148..473K}{\apjs}, \href{https://ui.adsabs.harvard.edu/abs/2003ApJS..148..473K}{148, 473}

\bibitem[{{Klinger} {et~al.}(2002){Klinger}, {Dickel}, {Fields}, \& {Milne}}]{2002AJ....124.2135K}
{Klinger}, R.~J., {Dickel}, J.~R., {Fields}, B.~D., and {Milne}, D.~K. 2002, \href{http://adsabs.harvard.edu/abs/2002AJ....124.2135K}{\aj}, \href{http://adsabs.harvard.edu/abs/2002AJ....124.2135K}{124, 2135}

\bibitem[{Klinger {et~al.}(2002)Klinger, Dickel, Fields, \& Milne}]{klinger2002peculiar}
Klinger, R.~J., Dickel, J.~R., Fields, B.~D., and Milne, D.~K. 2002, The Astronomical Journal, 124, 2135

\bibitem[{{Kothes} {et~al.}(2017){Kothes}, {Reich}, {Foster}, \& {Reich}}]{Kothes2017}
{Kothes}, R., {Reich}, P., {Foster}, T.~J., and {Reich}, W. 2017, \href{https://ui.adsabs.harvard.edu/abs/2017A&A...597A.116K}{A\&A}, \href{https://ui.adsabs.harvard.edu/abs/2017A&A...597A.116K}{597, A116}

\bibitem[{{Laki{\'c}evi{\'c}} {et~al.}(2015){Laki{\'c}evi{\'c}}, {van Loon}, {Meixner}, {Gordon}, {Bot}, {Roman-Duval}, {Babler}, {Bolatto}, {Engelbracht}, {Filipovi{\'c}}, {Hony}, {Indebetouw}, {Misselt}, {Montiel}, {Okumura}, {Panuzzo}, {Patat}, {Sauvage}, {Seale}, {Sonneborn}, {Temim}, {Uro{\v{s}}evi{\'c}}, \& {Zanardo}}]{2015ApJ...799...50L}
{Laki{\'c}evi{\'c}}, M., {van Loon}, J.~T., {Meixner}, M., {et~al.} 2015, \href{https://ui.adsabs.harvard.edu/abs/2015ApJ...799...50L}{\apj}, \href{https://ui.adsabs.harvard.edu/abs/2015ApJ...799...50L}{799, 50}

\bibitem[{{Lasker}(1977)}]{Lasker1977}
{Lasker}, B.~M. 1977, \href{https://ui.adsabs.harvard.edu/abs/1977PASP...89..474L}{\pasp}, \href{https://ui.adsabs.harvard.edu/abs/1977PASP...89..474L}{89, 474}

\bibitem[{{Lazarevi{\'c}} {et~al.}(2024{\natexlab{a}}){Lazarevi{\'c}}, {Filipovi{\'c}}, {Dai}, {Kothes}, {Ahmad}, {Alsaberi}, {Balzan}, {Barnes}, {Cotton}, {Edwards}, {Gordon}, {Haberl}, {Hopkins}, {Koribalski}, {Leahy}, {Maitra}, {Mi{\'c}i{\'c}}, {Rowell}, {Sasaki}, {Tothill}, {Umana}, \& {Velovi{\'c}}}]{2024PASA...41...32L}
{Lazarevi{\'c}}, S., {Filipovi{\'c}}, M.~D., {Dai}, S., {et~al.} 2024{\natexlab{a}}, \href{https://ui.adsabs.harvard.edu/abs/2024PASA...41...32L}{\pasa}, \href{https://ui.adsabs.harvard.edu/abs/2024PASA...41...32L}{41, e032}

\bibitem[{{Lazarevi{\'c}} {et~al.}(2024{\natexlab{b}}){Lazarevi{\'c}}, {Filipovi{\'c}}, {Koribalski}, {Smeaton}, {Hopkins}, {Alsaberi}, {Velovi{\'c}}, {Ball}, {Kothes}, {Leahy}, \& {Ingallinera}}]{Lazarevic2024}
{Lazarevi{\'c}}, S., {Filipovi{\'c}}, M.~D., {Koribalski}, B.~S., {et~al.} 2024{\natexlab{b}}, \href{https://ui.adsabs.harvard.edu/abs/2024RNAAS...8..107L}{Research Notes of the American Astronomical Society}, \href{https://ui.adsabs.harvard.edu/abs/2024RNAAS...8..107L}{8, 107}

\bibitem[{{Lazendic} {et~al.}(2000){Lazendic}, {Dickel}, {Haynes}, {Jones}, \& {White}}]{2000ApJ...540..808L}
{Lazendic}, J.~S., {Dickel}, J.~R., {Haynes}, R.~F., {Jones}, P.~A., and {White}, G.~L. 2000, \href{https://ui.adsabs.harvard.edu/abs/2000ApJ...540..808L}{\apj}, \href{https://ui.adsabs.harvard.edu/abs/2000ApJ...540..808L}{540, 808}

\bibitem[{{Leahy}(2017)}]{2017ApJ...837...36L}
{Leahy}, D.~A. 2017, \href{https://ui.adsabs.harvard.edu/abs/2017ApJ...837...36L}{\apj}, \href{https://ui.adsabs.harvard.edu/abs/2017ApJ...837...36L}{837, 36}

\bibitem[{{Leahy} {et~al.}(2025){Leahy}, {Ranasinghe}, {Hansen}, {Filipovi{\'c}}, \& {Smeaton}}]{2025PASP..137f4502L}
{Leahy}, D.~A., {Ranasinghe}, S., {Hansen}, J., {Filipovi{\'c}}, M.~D., and {Smeaton}, Z. 2025, \href{https://ui.adsabs.harvard.edu/abs/2025PASP..137f4502L}{\pasp}, \href{https://ui.adsabs.harvard.edu/abs/2025PASP..137f4502L}{137, 064502}

\bibitem[{{Long} {et~al.}(1981){Long}, {Helfand}, \& {Grabelsky}}]{Long1981}
{Long}, K.~S., {Helfand}, D.~J., and {Grabelsky}, D.~A. 1981, \href{https://ui.adsabs.harvard.edu/abs/1981ApJ...248..925L}{\apj}, \href{https://ui.adsabs.harvard.edu/abs/1981ApJ...248..925L}{248, 925}

\bibitem[{{Lopez} {et~al.}(2009){Lopez}, {Ramirez-Ruiz}, {Badenes}, {Huppenkothen}, {Jeltema}, \& {Pooley}}]{2009ApJ...706L.106L}
{Lopez}, L.~A., {Ramirez-Ruiz}, E., {Badenes}, C., {et~al.} 2009, \href{https://ui.adsabs.harvard.edu/abs/2009ApJ...706L.106L}{\apjl}, \href{https://ui.adsabs.harvard.edu/abs/2009ApJ...706L.106L}{706, L106}

\bibitem[{{Lopez} {et~al.}(2011){Lopez}, {Ramirez-Ruiz}, {Huppenkothen}, {Badenes}, \& {Pooley}}]{2011ApJ...732..114L}
{Lopez}, L.~A., {Ramirez-Ruiz}, E., {Huppenkothen}, D., {Badenes}, C., and {Pooley}, D.~A. 2011, \href{https://ui.adsabs.harvard.edu/abs/2011ApJ...732..114L}{\apj}, \href{https://ui.adsabs.harvard.edu/abs/2011ApJ...732..114L}{732, 114}

\bibitem[{{Luken} {et~al.}(2020){Luken}, {Filipovi{\'c}}, {Maxted}, {Kothes}, {Norris}, {Allison}, {Blackwell}, {Braiding}, {Brose}, {Burton}, {De Horta}, {Galvin}, {Harvey-Smith}, {Hurley-Walker}, {Leahy}, {Ralph}, {Roper}, {Rowell}, {Sushch}, {Uro{\v{s}}evi{\'c}}, \& {Wong}}]{Luken2020}
{Luken}, K.~J., {Filipovi{\'c}}, M.~D., {Maxted}, N.~I., {et~al.} 2020, \href{https://ui.adsabs.harvard.edu/abs/2020MNRAS.492.2606L}{MNRAS}, \href{https://ui.adsabs.harvard.edu/abs/2020MNRAS.492.2606L}{492, 2606}

\bibitem[{{Maggi} {et~al.}(2019){Maggi}, {Filipovi{\'c}}, {Vukoti{\'c}}, {Ballet}, {Haberl}, {Maitra}, {Kavanagh}, {Sasaki}, \& {Stupar}}]{Maggi2019}
{Maggi}, P., {Filipovi{\'c}}, M.~D., {Vukoti{\'c}}, B., {et~al.} 2019, \href{https://ui.adsabs.harvard.edu/abs/2019A&A...631A.127M}{\aap}, \href{https://ui.adsabs.harvard.edu/abs/2019A&A...631A.127M}{631, A127}

\bibitem[{{Maggi} {et~al.}(2016){Maggi}, {Haberl}, {Kavanagh}, {Sasaki}, {Bozzetto}, {Filipovi{\'c}}, {Vasilopoulos}, {Pietsch}, {Points}, {Chu}, {Dickel}, {Ehle}, {Williams}, \& {Greiner}}]{Maggi2016}
{Maggi}, P., {Haberl}, F., {Kavanagh}, P.~J., {et~al.} 2016, \href{https://ui.adsabs.harvard.edu/abs/2016A&A...585A.162M}{\aap}, \href{https://ui.adsabs.harvard.edu/abs/2016A&A...585A.162M}{585, A162}

\bibitem[{{Maitra} {et~al.}(2015){Maitra}, {Ballet}, {Filipovi{\'c}}, {Haberl}, {Tiengo}, {Grieve}, \& {Roper}}]{2015A&A...584A..41M}
{Maitra}, C., {Ballet}, J., {Filipovi{\'c}}, M.~D., {et~al.} 2015, \href{https://ui.adsabs.harvard.edu/abs/2015A&A...584A..41M}{\aap}, \href{https://ui.adsabs.harvard.edu/abs/2015A&A...584A..41M}{584, A41}

\bibitem[{{Maitra} {et~al.}(2021){Maitra}, {Esposito}, {Tiengo}, {Ballet}, {Haberl}, {Dai}, {Filipovi{\'c}}, \& {Pilia}}]{2021MNRAS.507L...1M}
{Maitra}, C., {Esposito}, P., {Tiengo}, A., {et~al.} 2021, \href{https://ui.adsabs.harvard.edu/abs/2021MNRAS.507L...1M}{\mnras}, \href{https://ui.adsabs.harvard.edu/abs/2021MNRAS.507L...1M}{507, L1}

\bibitem[{Mathewson and Clarke(1973)Mathewson \& Clarke}]{mathewson1973supernova}
Mathewson, D. and Clarke, J. 1973, The Astrophysical Journal, 180, 725

\bibitem[{{Mauch} {et~al.}(2003){Mauch}, {Murphy}, {Buttery}, {Curran}, {Hunstead}, {Piestrzynski}, {Robertson}, \& {Sadler}}]{2003MNRAS.342.1117M}
{Mauch}, T., {Murphy}, T., {Buttery}, H.~J., {et~al.} 2003, \href{https://ui.adsabs.harvard.edu/abs/2003MNRAS.342.1117M}{\mnras}, \href{https://ui.adsabs.harvard.edu/abs/2003MNRAS.342.1117M}{342, 1117}

\bibitem[{{McEntaffer} {et~al.}(2012){McEntaffer}, {Brantseg}, \& {Presley}}]{2012ApJ...756...17M}
{McEntaffer}, R.~L., {Brantseg}, T., and {Presley}, M. 2012, \href{https://ui.adsabs.harvard.edu/abs/2012ApJ...756...17M}{\apj}, \href{https://ui.adsabs.harvard.edu/abs/2012ApJ...756...17M}{756, 17}

\bibitem[{{Meixner} {et~al.}(2013){Meixner}, {Panuzzo}, {Roman-Duval}, {Engelbracht}, {Babler}, {Seale}, {Hony}, {Montiel}, {Sauvage}, {Gordon}, {Misselt}, {Okumura}, {Chanial}, {Beck}, {Bernard}, {Bolatto}, {Bot}, {Boyer}, {Carlson}, {Clayton}, {Chen}, {Cormier}, {Fukui}, {Galametz}, {Galliano}, {Hora}, {Hughes}, {Indebetouw}, {Israel}, {Kawamura}, {Kemper}, {Kim}, {Kwon}, {Lebouteiller}, {Li}, {Long}, {Madden}, {Matsuura}, {Muller}, {Oliveira}, {Onishi}, {Otsuka}, {Paradis}, {Poglitsch}, {Reach}, {Robitaille}, {Rubio}, {Sargent}, {Sewi{\l}o}, {Skibba}, {Smith}, {Srinivasan}, {Tielens}, {van Loon}, \& {Whitney}}]{Meixner2013}
{Meixner}, M., {Panuzzo}, P., {Roman-Duval}, J., {et~al.} 2013, \href{https://ui.adsabs.harvard.edu/abs/2013AJ....146...62M}{\aj}, \href{https://ui.adsabs.harvard.edu/abs/2013AJ....146...62M}{146, 62}

\bibitem[{{Mills} {et~al.}(1984){Mills}, {Turtle}, {Little}, \& {Durdin}}]{Mills1984}
{Mills}, B.~Y., {Turtle}, A.~J., {Little}, A.~G., and {Durdin}, J.~M. 1984, \href{https://ui.adsabs.harvard.edu/abs/1984AuJPh..37..321M}{Australian Journal of Physics}, \href{https://ui.adsabs.harvard.edu/abs/1984AuJPh..37..321M}{37, 321}

\bibitem[{{Milne} {et~al.}(1980){Milne}, {Caswell}, \& {Haynes}}]{Milne1980}
{Milne}, D.~K., {Caswell}, J.~L., and {Haynes}, R.~F. 1980, \href{https://ui.adsabs.harvard.edu/abs/1980MNRAS.191..469M}{\mnras}, \href{https://ui.adsabs.harvard.edu/abs/1980MNRAS.191..469M}{191, 469}

\bibitem[{{Owen} {et~al.}(2011){Owen}, {Filipovi{\'c}}, {Ballet}, {Haberl}, {Crawford}, {Payne}, {Sturm}, {Pietsch}, {Mereghetti}, {Ehle}, {Tiengo}, {Coe}, {Hatzidimitriou}, \& {Buckley}}]{2011A&A...530A.132O}
{Owen}, R.~A., {Filipovi{\'c}}, M.~D., {Ballet}, J., {et~al.} 2011, \href{https://ui.adsabs.harvard.edu/abs/2011A&A...530A.132O}{\aap}, \href{https://ui.adsabs.harvard.edu/abs/2011A&A...530A.132O}{530, A132}

\bibitem[{{Pavan} {et~al.}(2014{\natexlab{a}}){Pavan}, {Bordas}, {P{\"u}hlhofer}, {Filipovi{\'c}}, {De Horta}, {O'Brien}, {Balbo}, {Walter}, {Bozzo}, {Ferrigno}, {Crawford}, \& {Stella}}]{2014A&A...562A.122P}
{Pavan}, L., {Bordas}, P., {P{\"u}hlhofer}, G., {et~al.} 2014{\natexlab{a}}, \href{https://ui.adsabs.harvard.edu/abs/2014A&A...562A.122P}{\aap}, \href{https://ui.adsabs.harvard.edu/abs/2014A&A...562A.122P}{562, A122}

\bibitem[{{Pavan} {et~al.}(2014{\natexlab{b}}){Pavan}, {Bordas}, {P{\"u}hlhofer}, {Filipovic}, {de Horta}, {O'Brien}, {Crawford}, {Balbo}, {Walter}, {Bozzo}, {Ferrigno}, \& {Stella}}]{2014IJMPS..2860172P}
{Pavan}, L., {Bordas}, P., {P{\"u}hlhofer}, G., {et~al.} 2014{\natexlab{b}}, in International Journal of Modern Physics Conference Series, Vol.~28, International Journal of Modern Physics Conference Series, 1460172

\bibitem[{{Pavan} {et~al.}(2016){Pavan}, {P{\"u}hlhofer}, {Bordas}, {Audard}, {Balbo}, {Bozzo}, {Eckert}, {Ferrigno}, {Filipovi{\'c}}, {Verdugo}, \& {Walter}}]{2016A&A...591A..91P}
{Pavan}, L., {P{\"u}hlhofer}, G., {Bordas}, P., {et~al.} 2016, \href{https://ui.adsabs.harvard.edu/abs/2016A&A...591A..91P}{\aap}, \href{https://ui.adsabs.harvard.edu/abs/2016A&A...591A..91P}{591, A91}

\bibitem[{{Pavlovi{\'c}} {et~al.}(2018){Pavlovi{\'c}}, {Uro{\v{s}}evi{\'c}}, {Arbutina}, {Orlando}, {Maxted}, \& {Filipovi{\'c}}}]{Pavlovic2018}
{Pavlovi{\'c}}, M.~Z., {Uro{\v{s}}evi{\'c}}, D., {Arbutina}, B., {et~al.} 2018, \href{https://ui.adsabs.harvard.edu/abs/2018ApJ...852...84P}{Astrophys. J.}, \href{https://ui.adsabs.harvard.edu/abs/2018ApJ...852...84P}{852, 84}

\bibitem[{{Payne} {et~al.}(2008){Payne}, {White}, \& {Filipovi{\'c}}}]{Payne2008}
{Payne}, J.~L., {White}, G.~L., and {Filipovi{\'c}}, M.~D. 2008, \href{https://ui.adsabs.harvard.edu/abs/2008MNRAS.383.1175P}{\mnras}, \href{https://ui.adsabs.harvard.edu/abs/2008MNRAS.383.1175P}{383, 1175}

\bibitem[{{Payne} {et~al.}(2007){Payne}, {White}, {Filipovi{\'c}}, \& {Pannuti}}]{Payne2007}
{Payne}, J.~L., {White}, G.~L., {Filipovi{\'c}}, M.~D., and {Pannuti}, T.~G. 2007, \href{https://ui.adsabs.harvard.edu/abs/2007MNRAS.376.1793P}{\mnras}, \href{https://ui.adsabs.harvard.edu/abs/2007MNRAS.376.1793P}{376, 1793}

\bibitem[{{Pennock} {et~al.}(2021){Pennock}, {van Loon}, {Filipovi{\'c}}, {Andernach}, {Haberl}, {Kothes}, {Lenc}, {Rudnick}, {White}, {Agliozzo}, {Ant{\'o}n}, {Boji{\v{c}}i{\'c}}, {Bomans}, {Collier}, {Crawford}, {Hopkins}, {Jeganathan}, {Kavanagh}, {Koribalski}, {Leahy}, {Maggi}, {Maitra}, {Marvil}, {Micha{\l}owski}, {Norris}, {Oliveira}, {Payne}, {Sano}, {Sasaki}, {Staveley-Smith}, \& {Vardoulaki}}]{2021MNRAS.506.3540P}
{Pennock}, C.~M., {van Loon}, J.~T., {Filipovi{\'c}}, M.~D., {et~al.} 2021, \href{https://ui.adsabs.harvard.edu/abs/2021MNRAS.506.3540P}{\mnras}, \href{https://ui.adsabs.harvard.edu/abs/2021MNRAS.506.3540P}{506, 3540}

\bibitem[{{Pietrzy{\'n}ski} {et~al.}(2019){Pietrzy{\'n}ski}, {Graczyk}, {Gallenne}, {Gieren}, {Thompson}, {Pilecki}, {Karczmarek}, {G{\'o}rski}, {Suchomska}, {Taormina}, {Zgirski}, {Wielg{\'o}rski}, {Ko{\l}aczkowski}, {Konorski}, {Villanova}, {Nardetto}, {Kervella}, {Bresolin}, {Kudritzki}, {Storm}, {Smolec}, \& {Narloch}}]{Pietrzynski2019}
{Pietrzy{\'n}ski}, G., {Graczyk}, D., {Gallenne}, A., {et~al.} 2019, \href{http://adsabs.harvard.edu/abs/2019Natur.567..200P}{Nature}, \href{http://adsabs.harvard.edu/abs/2019Natur.567..200P}{567, 200}

\bibitem[{{Points} {et~al.}(2024){Points}, {Long}, {Blair}, {Williams}, {Chu}, {Winkler}, {White}, {Rest}, {Li}, \& {Valdes}}]{2024ApJ...974...70P}
{Points}, S.~D., {Long}, K.~S., {Blair}, W.~P., {et~al.} 2024, \href{https://ui.adsabs.harvard.edu/abs/2024ApJ...974...70P}{\apj}, \href{https://ui.adsabs.harvard.edu/abs/2024ApJ...974...70P}{974, 70}

\bibitem[{{Porth} {et~al.}(2013){Porth}, {Komissarov}, \& {Keppens}}]{2013MNRAS.431L..48P}
{Porth}, O., {Komissarov}, S.~S., and {Keppens}, R. 2013, \href{https://ui.adsabs.harvard.edu/abs/2013MNRAS.431L..48P}{\mnras}, \href{https://ui.adsabs.harvard.edu/abs/2013MNRAS.431L..48P}{431, L48}

\bibitem[{{Ranasinghe} and {Leahy}(2019{\natexlab{a}}){Ranasinghe} \& {Leahy}}]{Multipole2019}
{Ranasinghe}, S. and {Leahy}, D. 2019{\natexlab{a}}, {Journal of High Energy Physics, Gravitation and Cosmology, \href{https://ui.adsabs.harvard.edu/abs/2019arXiv190911803R}{}\eprint  \href{https://doi.org/10.4236/jhepgc.2019.53046}{(DOI:  10.4236/jhepgc.2019.53046)}, 907}

\bibitem[{{Ranasinghe} and {Leahy}(2019{\natexlab{b}}){Ranasinghe} \& {Leahy}}]{2019arXiv190911803R}
{Ranasinghe}, S. and {Leahy}, D. 2019{\natexlab{b}}, {arXiv e-prints, \href{https://ui.adsabs.harvard.edu/abs/2019arXiv190911803R}{}\eprint  \href{https://arxiv.org/abs/1909.11803}{(DOI:  10.48550/arXiv.1909.11803)}, arXiv:1909.11803}

\bibitem[{{Reid} {et~al.}(2015){Reid}, {Stupar}, {Bozzetto}, {Parker}, \& {Filipovi{\'c}}}]{2015MNRAS.454..991R}
{Reid}, W.~A., {Stupar}, M., {Bozzetto}, L.~M., {Parker}, Q.~A., and {Filipovi{\'c}}, M.~D. 2015, \href{https://ui.adsabs.harvard.edu/abs/2015MNRAS.454..991R}{\mnras}, \href{https://ui.adsabs.harvard.edu/abs/2015MNRAS.454..991R}{454, 991}

\bibitem[{{Reynolds} {et~al.}(2017){Reynolds}, {Pavlov}, {Kargaltsev}, {Klingler}, {Renaud}, \& {Mereghetti}}]{2017Reynolds}
{Reynolds}, S.~P., {Pavlov}, G.~G., {Kargaltsev}, O., {et~al.} 2017, \href{https://ui.adsabs.harvard.edu/abs/2017SSRv..207..175R}{\ssr}, \href{https://ui.adsabs.harvard.edu/abs/2017SSRv..207..175R}{207, 175}

\bibitem[{Rho and Petre(1998)Rho \& Petre}]{rho1998mixed}
Rho, J. and Petre, R. 1998, The Astrophysical Journal Letters, 503, L167

\bibitem[{{Sano} {et~al.}(2021){Sano}, {Suzuki}, {Nobukawa}, {Filipovi{\'c}}, {Fukui}, \& {Moriya}}]{2021ApJ...923...15S}
{Sano}, H., {Suzuki}, H., {Nobukawa}, K.~K., {et~al.} 2021, \href{https://ui.adsabs.harvard.edu/abs/2021ApJ...923...15S}{\apj}, \href{https://ui.adsabs.harvard.edu/abs/2021ApJ...923...15S}{923, 15}

\bibitem[{{Sano} {et~al.}(2022){Sano}, {Yamaguchi}, {Aruga}, {Fukui}, {Tachihara}, {Filipovi{\'c}}, \& {Rowell}}]{2022ApJ...933..157S}
{Sano}, H., {Yamaguchi}, H., {Aruga}, M., {et~al.} 2022, \href{https://ui.adsabs.harvard.edu/abs/2022ApJ...933..157S}{\apj}, \href{https://ui.adsabs.harvard.edu/abs/2022ApJ...933..157S}{933, 157}

\bibitem[{{Sasaki} {et~al.}(2025){Sasaki}, {Zangrandi}, {Filipovi{\'c}}, {Alsaberi}, {Collier}, {Haberl}, {Heywood}, {Kavanagh}, {Koribalski}, {Kothes}, {Lazarevi{\'c}}, {Maggi}, {Maitra}, {Points}, {Smeaton}, \& {Velovi{\'c}}}]{2025A&A...693L..15S}
{Sasaki}, M., {Zangrandi}, F., {Filipovi{\'c}}, M., {et~al.} 2025, \href{https://ui.adsabs.harvard.edu/abs/2025A&A...693L..15S}{\aap}, \href{https://ui.adsabs.harvard.edu/abs/2025A&A...693L..15S}{693, L15}

\bibitem[{{Sault} {et~al.}(1995){Sault}, {Teuben}, \& {Wright}}]{1995ASPC...77..433S}
{Sault}, R.~J., {Teuben}, P.~J., and {Wright}, M.~C.~H. 1995, in Astronomical Society of the Pacific Conference Series, Vol.~77, Astronomical Data Analysis Software and Systems IV, ed. R.~A. {Shaw}, H.~E. {Payne}, and J.~J.~E. {Hayes}, 433

\bibitem[{{Smeaton} {et~al.}(2025){Smeaton}, {Filipovi{\'c}}, {Alsaberi}, {Arbutina}, {Cotton}, {Crawford}, {Hopkins}, {Kothes}, {Leahy}, {Payne}, {Rajabpour}, {Sano}, {Sasaki}, {Uro{\v{s}}evi{\'c}}, \& {van Loon}}]{2025arXiv250615067S}
{Smeaton}, Z.~J., {Filipovi{\'c}}, M.~D., {Alsaberi}, R.~Z.~E., {et~al.} 2025, {arXiv e-prints, \href{https://ui.adsabs.harvard.edu/abs/2025arXiv250615067S}{}\eprint  \href{https://doi.org/10.48550/arXiv.2506.15067}{(DOI:  10.48550/arXiv.2506.15067)}, arXiv:2506.15067}

\bibitem[{{Smeaton} {et~al.}(2024{\natexlab{a}}){Smeaton}, {Filipovi{\'c}}, {Koribalski}, {Lazarevi{\'c}}, {Alsaberi}, {Becker}, {Dage}, {Gordon}, {Hopkins}, {Kothes}, {Leahy}, \& {Mitras̆inovi{\'c}}}]{Smeaton2024}
{Smeaton}, Z.~J., {Filipovi{\'c}}, M.~D., {Koribalski}, B.~S., {et~al.} 2024{\natexlab{a}}, \href{https://ui.adsabs.harvard.edu/abs/2024RNAAS...8..158S}{Research Notes of the American Astronomical Society}, \href{https://ui.adsabs.harvard.edu/abs/2024RNAAS...8..158S}{8, 158}

\bibitem[{{Smeaton} {et~al.}(2024{\natexlab{b}}){Smeaton}, {Filipovi{\'c}}, {Lazarevi{\'c}}, {Alsaberi}, {Ahmad}, {Araya}, {Ball}, {Bordiu}, {Buemi}, {Bufano}, {Dai}, {Haberl}, {Hopkins}, {Ingallinera}, {Jarrett}, {Koribalski}, {Kothes}, {Kraan-Korteweg}, {Leahy}, {Lundqvist}, {Maitra}, {Martin}, {Payne}, {Rowell}, {Sano}, {Sasaki}, {Soria}, {Steyn}, {Umana}, {Uro{\v{s}}evi{\'c}}, {Velovi{\'c}}, {Vernstrom}, {Vukoti{\'c}}, \& {West}}]{Perun}
{Smeaton}, Z.~J., {Filipovi{\'c}}, M.~D., {Lazarevi{\'c}}, S., {et~al.} 2024{\natexlab{b}}, \href{https://ui.adsabs.harvard.edu/abs/2024MNRAS.534.2918S}{\mnras}, \href{https://ui.adsabs.harvard.edu/abs/2024MNRAS.534.2918S}{534, 2918}

\bibitem[{{Smith} and {MCELS Team}(1999){Smith} \& {MCELS Team}}]{1999IAUS..190...28S}
{Smith}, R.~C. and {MCELS Team}. 1999, in IAU Symposium, Vol. 190, New Views of the Magellanic Clouds, ed. Y.~H. {Chu}, N.~{Suntzeff}, J.~{Hesser}, and D.~{Bohlender}, 28

\bibitem[{{Tingay} {et~al.}(2013){Tingay}, {Goeke}, {Bowman}, {Emrich}, {Ord}, {Mitchell}, {Morales}, {Booler}, {Crosse}, {Wayth}, {Lonsdale}, {Tremblay}, {Pallot}, {Colegate}, {Wicenec}, {Kudryavtseva}, {Arcus}, {Barnes}, {Bernardi}, {Briggs}, {Burns}, {Bunton}, {Cappallo}, {Corey}, {Deshpande}, {Desouza}, {Gaensler}, {Greenhill}, {Hall}, {Hazelton}, {Herne}, {Hewitt}, {Johnston-Hollitt}, {Kaplan}, {Kasper}, {Kincaid}, {Koenig}, {Kratzenberg}, {Lynch}, {Mckinley}, {Mcwhirter}, {Morgan}, {Oberoi}, {Pathikulangara}, {Prabu}, {Remillard}, {Rogers}, {Roshi}, {Salah}, {Sault}, {Udaya-Shankar}, {Schlagenhaufer}, {Srivani}, {Stevens}, {Subrahmanyan}, {Waterson}, {Webster}, {Whitney}, {Williams}, {Williams}, \& {Wyithe}}]{Tingay2013}
{Tingay}, S.~J., {Goeke}, R., {Bowman}, J.~D., {et~al.} 2013, \href{https://ui.adsabs.harvard.edu/abs/2013PASA...30....7T}{\pasa}, \href{https://ui.adsabs.harvard.edu/abs/2013PASA...30....7T}{30, e007}

\bibitem[{{Uro{\v{s}}evi{\'c}} {et~al.}(2018){Uro{\v{s}}evi{\'c}}, {Pavlovi{\'c}}, \& {Arbutina}}]{2018ApJ...855...59U}
{Uro{\v{s}}evi{\'c}}, D., {Pavlovi{\'c}}, M.~Z., and {Arbutina}, B. 2018, \href{https://ui.adsabs.harvard.edu/abs/2018ApJ...855...59U}{\apj}, \href{https://ui.adsabs.harvard.edu/abs/2018ApJ...855...59U}{855, 59}

\bibitem[{{Velovi{\'c}} {et~al.}(2023){Velovi{\'c}}, {Cotton}, {Filipovi{\'c}}, {Norris}, {Barnes}, \& {Condon}}]{2023MNRAS.523.1933V}
{Velovi{\'c}}, V., {Cotton}, W.~D., {Filipovi{\'c}}, M.~D., {et~al.} 2023, \href{https://ui.adsabs.harvard.edu/abs/2023MNRAS.523.1933V}{\mnras}, \href{https://ui.adsabs.harvard.edu/abs/2023MNRAS.523.1933V}{523, 1933}

\bibitem[{{Velovi{\'c}} {et~al.}(2022){Velovi{\'c}}, {Filipovi{\'c}}, {Barnes}, {Norris}, {Tremblay}, {Heald}, {Rudnick}, {Shabala}, {Pannuti}, {Andernach}, {Titov}, {Waddell}, {Koribalski}, {Grupe}, {Jarrett}, {Alsaberi}, {Carretti}, {Collier}, {Einecke}, {Galvin}, {Hotan}, {Manojlovi{\'c}}, {Marvil}, {Nandra}, {Reiprich}, {Rowell}, {Salvato}, \& {Whiting}}]{2022MNRAS.516.1865V}
{Velovi{\'c}}, V., {Filipovi{\'c}}, M.~D., {Barnes}, L., {et~al.} 2022, \href{https://ui.adsabs.harvard.edu/abs/2022MNRAS.516.1865V}{\mnras}, \href{https://ui.adsabs.harvard.edu/abs/2022MNRAS.516.1865V}{516, 1865}

\bibitem[{Virtanen {et~al.}(2020)Virtanen, Gommers, Oliphant, Haberland, Reddy, Cournapeau, Burovski, Peterson, Weckesser, Bright, {van der Walt}, Brett, Wilson, Millman, Mayorov, Nelson, Jones, Kern, Larson, Carey, Polat, Feng, Moore, {VanderPlas}, Laxalde, Perktold, Cimrman, Henriksen, Quintero, Harris, Archibald, Ribeiro, Pedregosa, {van Mulbregt}, \& {SciPy 1.0 Contributors}}]{Virtanen2020}
Virtanen, P., Gommers, R., Oliphant, T.~E., {et~al.} 2020, \href{https://rdcu.be/b08Wh}{Nature Methods}, \href{https://rdcu.be/b08Wh}{17, 261}

\bibitem[{{Wayth} {et~al.}(2015){Wayth}, {Lenc}, {Bell}, {Callingham}, {Dwarakanath}, {Franzen}, {For}, {Gaensler}, {Hancock}, {Hindson}, {Hurley-Walker}, {Jackson}, {Johnston-Hollitt}, {Kapi{\'n}ska}, {McKinley}, {Morgan}, {Offringa}, {Procopio}, {Staveley-Smith}, {Wu}, {Zheng}, {Trott}, {Bernardi}, {Bowman}, {Briggs}, {Cappallo}, {Corey}, {Deshpande}, {Emrich}, {Goeke}, {Greenhill}, {Hazelton}, {Kaplan}, {Kasper}, {Kratzenberg}, {Lonsdale}, {Lynch}, {McWhirter}, {Mitchell}, {Morales}, {Morgan}, {Oberoi}, {Ord}, {Prabu}, {Rogers}, {Roshi}, {Shankar}, {Srivani}, {Subrahmanyan}, {Tingay}, {Waterson}, {Webster}, {Whitney}, {Williams}, \& {Williams}}]{Wayth2015}
{Wayth}, R.~B., {Lenc}, E., {Bell}, M.~E., {et~al.} 2015, \href{https://ui.adsabs.harvard.edu/abs/2015PASA...32...25W}{\pasa}, \href{https://ui.adsabs.harvard.edu/abs/2015PASA...32...25W}{32, e025}

\bibitem[{{Williams} and {Chu}(2005){Williams} \& {Chu}}]{2005ApJ...635.1077W}
{Williams}, R.~M. and {Chu}, Y.~H. 2005, \href{https://ui.adsabs.harvard.edu/abs/2005ApJ...635.1077W}{\apj}, \href{https://ui.adsabs.harvard.edu/abs/2005ApJ...635.1077W}{635, 1077}

\bibitem[{Williams {et~al.}(2005)Williams, Chu, Dickel, Gruendl, Seward, Guerrero, \& Hobbs}]{williams2005supernova}
Williams, R.~M., Chu, Y.-H., Dickel, J., {et~al.} 2005, The Astrophysical Journal, 628, 704

\bibitem[{{Williams} {et~al.}(1999){Williams}, {Chu}, {Dickel}, {Petre}, {Smith}, \& {Tavarez}}]{Williams1999}
{Williams}, R.~M., {Chu}, Y.-H., {Dickel}, J.~R., {et~al.} 1999, \href{https://ui.adsabs.harvard.edu/abs/1999ApJS..123..467W}{\apjs}, \href{https://ui.adsabs.harvard.edu/abs/1999ApJS..123..467W}{123, 467}

\bibitem[{{Xiao} {et~al.}(2008){Xiao}, {F{\"u}rst}, {Reich}, \& {Han}}]{Xiao2008}
{Xiao}, L., {F{\"u}rst}, E., {Reich}, W., and {Han}, J.~L. 2008, \href{https://ui.adsabs.harvard.edu/abs/2008A&A...482..783X}{\aap}, \href{https://ui.adsabs.harvard.edu/abs/2008A&A...482..783X}{482, 783}

\bibitem[{{Yew} {et~al.}(2021){Yew}, {Filipovi{\'c}}, {Stupar}, {Points}, {Sasaki}, {Maggi}, {Haberl}, {Kavanagh}, {Parker}, {Crawford}, {Vukoti{\'c}}, {Uro{\v{s}}evi{\'c}}, {Sano}, {Seitenzahl}, {Rowell}, {Leahy}, {Bozzetto}, {Maitra}, {Leverenz}, {Payne}, {Park}, {Alsaberi}, \& {Pannuti}}]{Yew2021}
{Yew}, M., {Filipovi{\'c}}, M.~D., {Stupar}, M., {et~al.} 2021, \href{https://ui.adsabs.harvard.edu/abs/2021MNRAS.500.2336Y}{\mnras}, \href{https://ui.adsabs.harvard.edu/abs/2021MNRAS.500.2336Y}{500, 2336}

\bibitem[{{Yusef-Zadeh} and {Bally}(1987){Yusef-Zadeh} \& {Bally}}]{1987Natur.330..455Y}
{Yusef-Zadeh}, F. and {Bally}, J. 1987, \href{https://ui.adsabs.harvard.edu/abs/1987Natur.330..455Y}{\nat}, \href{https://ui.adsabs.harvard.edu/abs/1987Natur.330..455Y}{330, 455}

\bibitem[{{Zangrandi} {et~al.}(2024){Zangrandi}, {Jurk}, {Sasaki}, {Knies}, {Filipovi{\'c}}, {Haberl}, {Kavanagh}, {Maitra}, {Maggi}, {Saeedi}, {Bernreuther}, {Koribalski}, {Points}, \& {Staveley-Smith}}]{Zangrandi2024}
{Zangrandi}, F., {Jurk}, K., {Sasaki}, M., {et~al.} 2024, \href{https://ui.adsabs.harvard.edu/abs/2024A&A...692A.237Z}{\aap}, \href{https://ui.adsabs.harvard.edu/abs/2024A&A...692A.237Z}{692, A237}

\end{thebibliography}
}

\clearpage

{\ }

\clearpage

{\ }

\newpage

\begin{strip}

{\ }
\vskip-40mm


\naslov{Multifrekventna radio-kontinum studija ostatka supernove u Velikom Magelanovom oblaku {\rm N}206 (Kozje Oko) i ``cik-cak'' maglina pulsarskog vetra}


\authors{
M. Ghavam$^1$,
Z. J. Smeaton$^1$,
{\rbf M. D. Filipovi{\cc}$^1$},
R. Z. E. Alsaberi$^{2,1}$,
C. Bordiu$^3$,
}
\authors{
W. D. Cotton$^{4, 5}$,
E. J. Crawford$^{1}$,
A. M. Hopkins$^6$, 
R. Kothes$^7$,
{\rbf S. Lazarevi{\cc}$^{1}$},
}
\authors{
D. Leahy$^8$,
N. Rajabpour$^1$,
S. Ranasinghe$^8$,
G. P. Rowell$^9$,
H. Sano$^{2}$,
M. Sasaki$^{10}$, 
}
\authors{
D. Shobhana$^{1}$,
K. Tsuge$^{2,11}$,
{\rbf D. Uro{\ss}evi{\cc}$^{12}$ i}
N. F. H. Tothill$^{1}$
}

\vskip3mm



\address{$^1$Western Sydney University, Locked Bag 1797, Penrith South DC, NSW 2751, Australia}
\address{$^2$Faculty of Engineering, Gifu University, 1-1 Yanagido, Gifu 501-1193, Japan}
\address{$^3$INAF-Osservatorio Astrofisico di Catania, Via Santa Sofía 78, I-95123 Catania, Italy}
\address{$^4$National Radio Astronomy Observatory, 520 Edgemont Road, Charlottesville, VA 22903, USA}
\address{$^5$South African Radio Astronomy Observatory Liesbeek House, River Park, Gloucester Road Cape Town, 7700, South Africa}
\address{$^6$School of Mathematical and Physical Sciences, 12 Wally’s Walk, Macquarie University, NSW 2109, Australia}
\address{$^7$Dominion Radio Astrophysical Observatory, Herzberg Astronomy \& Astrophysics, National Research Council Canada, P.O. Box 248, Penticton}
\address{$^8$Department of Physics and Astronomy, University of Calgary, Calgary, Alberta, T2N IN4, Canada}
\address{$^9$School of Physics, Chemistry and Earth Sciences, The University of Adelaide, Adelaide, 5005, Australia}
\address{$^{10}$Dr Karl Remeis Observatory, Erlangen Centre for Astroparticle Physics, Friedrich-Alexander-Universit\"{a}t Erlangen-N\"{u}rnberg, Sternwartstra{\ss}e 7, 96049 Bamberg, Germany}
\address{$^{11}$Institute for Advanced Study, Gifu University, 1-1 Yanagido, Gifu 501-1193, Japan}
\address{$^{12}${\rit Katedra za Astronomiju, Matematichki Fakultet, Univer{z}iteta u Beogradu, Studen{t}{s}ki trg 16, 11000 Beograd, Srbija}}

\Email{19594271@student.westernsydney.edu.au}

\vskip3mm


\centerline{{\rrm UDK} \udc}


\vskip1mm

\centerline{\rit Uredjivaqki prilog}

\vskip.7cm

\baselineskip=3.8truemm

\begin{multicols}{2}

{
\rrm

Predstavljamo nova radio-kontinum posmatranja ostatka supernove (OSN) {\rm N}206 u Velikom Magelanovom oblaku (VMO) kome smo dali nadimak ``Kozje Oko''. Kozje Oko sadr{\zz}i unutra{\ss}nju radio strukturu koja je verovatno maglina pulsarskog vetra (MPV), koju detaljno analiziramo. Koristimo nova radio-kontinum posmatranja sa Australijskog Teleskopog Kompaktnog Niza (ATKN) teleskopa, kao i nekoliko arhivskih radio posmatranja za izraqunavanje spektralnog indeksa. Na{\ss}li smo strmi spektralni indeks za ceo OSN ($\alpha = -0.60\pm0.02$), i ravan spektralni indeks za MPV ($\alpha = -0.16\pm0.03$). Tako{\dj}e merimo polarizaciju i svojstva magnetnog polja MPV. Ranije predlo{\zz}eno kao linearna struktura, nova posmatranja pokazuju neobiqnu ``cik-cak'' strukturu, vidljivu ekskluzivno u radio-kontinumu ukljuquju{\cc}i i linearnu polarizaciju, i orijentaciju magnetnog polja. Poreklo ove ``cik-cak'' strukture je i dalje nejasno, ali predla{\zz}emo jedan od mogu{\cc}ih scenarija koji zahteva nova posmatranja.

{\ }

}

\end{multicols}

\end{strip}


\end{document}